\documentclass[fleqn,usenatbib]{mnras}

\usepackage{newtxtext,newtxmath}

\usepackage[T1]{fontenc}
\DeclareRobustCommand{\VAN}[3]{#2}
\let\VANthebibliography\thebibliography
\def\thebibliography{\DeclareRobustCommand{\VAN}[3]{##3}\VANthebibliography}

\usepackage{graphicx}	
\graphicspath{ {./figures/} }
\usepackage{amsmath}	

\title[AGN and star formation]{AGN and Star Formation feedback in the evolution of galaxy outflows}

\author[Clavijo-Bohórquez, de Gouveia Dal Pino and Melioli]{
William E. Clavijo-Bohórquez$^{1}$\thanks{E-mail: weclavijob@usp.br (WECB)},
Elisabete M. de Gouveia Dal Pino 
$^{2}$\thanks{E-mail: dalpino@iag.usp.br (EMdGDP)},
Claudio Melioli$^{3}$\thanks{E-mail: claudio.melioli@unimore.it (CM)}
\\
$^{1}$Universidade de S\~{a}o Paulo, Instituto de Física, R. do Matão, 1371 - São Paulo - SP, 05508-090, Brasil\\
$^{2}$ Universidade de S\~{a}o Paulo, Instituto de Astronomia, Geof\'{i}sica e Ci\^{e}ncias Atmosf\'{e}ricas, Departamento de Astronomia, \\
$\;\;\;$1226 R. do Mat\~{a}o, S\~{a}o Paulo, 05508-090, Brasil\\
$^{3}$Universisty of Modena and Regio Emilia, Dipartamento di Scienze e Metodi dell'Ingengneria - DISMI, Via Amendola 2, 42122 Reggio Emilia, Italy
}

\date{Accepted XXX. Received YYY; in original form ZZZ}

\pubyear{2015}

\begin{document}
\label{firstpage}
\pagerange{\pageref{firstpage}--\pageref{lastpage}}
\maketitle

\begin{abstract}
We conducted 3D-MHD simulations to investigate the feedback processes in the central 1-kpc scale of galaxies hosting both active star formation (SF)  and an  AGN wind. Our simulations naturally generated a turbulent and clumpy interstellar medium driven by SF evolution. We found that the AGN wind duty cycle plays a crucial role in shaping the evolution of the outflows. This cycle consists of an active, a remnant and an inactive phase, lasting up to 1.5 Myr. The duration of the cycle increases with larger star formation rate (SFR) and smaller AGN wind power (tested for luminosities $10^{42} - 10^{44}$ erg s$^{-1}$ and SFR$=$1 – 1000 M$_{\odot}$ yr$^{-1}$). The feedback on SF, whether positive or negative, depends on various factors, including the AGN outflow opening angle, power, and phase of activity, as well as the initial SFR. The passage of the AGN wind enhances SF in a ring around it, resembling the structures observed in ULIRGs, and is stronger for larger AGN power or SFR. Also, a higher SFR enhances the mixing of interstellar matter with the AGN wind, resulting in a greater number of colder, denser structures with volume filling factors $\sim$ 0.02 to 0.12 and velocities comparable to those observed in Seyferts and LINERs, but smaller than those observed in ULIRGs. The efficiency of the AGN wind in transporting mass to kiloparsec distances diminishes with increasing SFR. The mass loss rates range from 50 to 250 M$_{\odot}$ yr$^{-1}$ within the initial 2 Myr of evolution, which aligns with observed rates in nearby Seyferts and ULIRGs. 
\end{abstract}




\section{Introduction} \label{sec:intro}
The impact of  supermassive black holes (SMBHs) on the evolution of their host  galaxies is known as feedback. The (momentum-driven) jets from active galaxies (AGN) are generally believed to be the main agents to prevent the cooling of the gas at the large (dozen-kpc)  scales where these jets impinge into the warm and hot intergalactic medium (IGM) (e.g. \citealt{SilkRees1998},  \citealt{Netzer2015}, \citealt{Falceta-Goncalves2010}). This process, often denominated "maintenance mode", is complemented at smaller (sub-kpc) scales, by the so called  “quasar mode” where (radiation-driven) outflows are believed to be responsible for sweeping the gas from the galaxy inner radii (e.g., \citealt{McNamara_2012}; \citealt{Morganti2021a}; \citealt{Morganti2021}) and even quench star formation  (\citealt{Nesvadba2010.521}, \citealt{Wylezalek2016.461}, \citealt{Mukherjee2018a}). On the other hand, with the recent increase of higher resolution observations revealing multiple phase structures of  gas in different classes of AGN, the aforementioned picture of AGN feedback has become more complex (e.g., \citealt{harrison2020proc}, \citealt{Scholtz2020.492MNRAS}, \citealt{Morganti2021},  \citealt{Aalto2020}, \citealt{Pereira-Santaella2018a}). The contribution in various classes of AGN to the feedback and the scales at which they operate are not yet completely understood (e.g. \citealt{Murthy2022NatAs...6..488M}). For instance, it is not clear yet whether galactic winds driven by star formation (SF) (also denominated supernova-driven winds) play some role in the feedback process too, blurring the effects of the AGN winds (e.g. \citealt{MelioliDalPino2015ApJ...812...90M}, \citealt{DeGouveiaDalPino2018proc}, \citealt{SilkRees1998}, \citealt{Netzer2015}, \citealt{Harrison2018a}, \citealt{Gowardhan2018b}, \citealt{Wylezalek2018}, \citealt{Pereira-Santaella2021}). At the same time, cosmological simulations, specially those focused on the  formation of SMBHs  (e.g. \citealt{Barai_dGDP2019}) or the production of cosmic rays and diffuse emission  \citep[e.g.][]{Hussain_etal2023}, have demanded more realistic constraints on the processes of energy transfer from the AGN and SB winds to the interstellar medium (ISM) and IGM.

The gaseous outflows observed in a high variety of AGN (e.g. \citealt{Tadhunter2008}; \citealt{Harrison2018a}, \citealt{Veilleux2020}, \citealt{Morganti2015}), are often limited to the central kiloparsec or sub-kiloparsec regions of the galaxy and carry relatively small amounts of gas  (see, e.g., \citealt{Holt2008}, \citealt{Rose2017....474...128-156, Rose2018a},  \citealt{Oosterloo2017...608...A38}, \citealt{Oosterloo2019A&A...632A..66O},  \citealt{Baron2018d}, \citealt{Bischetti2019...628...A118}, \citealt{Scholtz2020.492MNRAS},  \citealt{Santoro2018...617...A139, Santoro2020}). Their impact effects may change during the various stages in the life of the AGN (e.g. \citealt{Morganti2021}). 
Previous numerical studies have indicated that the compression by the jet-driven shocks can enhance star formation in the disk, but the enhanced turbulent dispersion in the disk can also lead to quenching of star formation, as indicated by observations. Whether positive or negative feedback prevails depends on jet power, ISM density, jet orientation with respect to the disk, and the time-scale under consideration \citep{Sutherland2007,  Wagner_2012, Mukherjee2016, MelioliDalPino2015ApJ...812...90M, Mukherjee2018a,Mukherjee2018c}, \citealt{Tanner_2022}.

In this work, in order to improve our understanding on the formation of these outflows and on how they impact on the galaxy evolution at the inner regions in different  stages of the AGN activity, we  present three-dimensional (3D) MHD simulations of the sub-kpc-scale core region of active galaxies with intense star-formation. Distinctly from previous works that have simulated the interaction of an AGN jet with a clumpy ISM with large filling factor fractal distribution of high-density clouds (e.g. \citealt{Wagner2011}, \citealt{Wagner_2012}, \citealt{Mukherjee2016}, \citealt{Mukherjee2018a, Mukherjee2018c}, \citealt{Murthy2022NatAs...6..488M}, \citealt{Tanner_2022}), we here consider fiducial values of star formation and supernova rates in an initially smooth multi-phase galactic disk in order to  allow for natural formation of a turbulent and  structured ISM where the AGN jet is then injected. A galactic wind driven by star formation is also naturally produced in this environment and we explore the interactions and relative role of both the AGN-driven and the SF-driven outflows on the feedback processes, including the temporary suppression of gas infall due to the AGN outflow. We also explore the effects of initial conditions, by launching the AGN outflow in different stages of the galactic ISM evolution. 

This work extends upon previous one by \cite{MelioliDalPino2015ApJ...812...90M}, where the combined effects of SF and AGN outflows were explored in the context of Seyfert galaxies and the main conclusion was that the SF is the main responsible for the mass loading in the outflows. The AGN jet alone would be unable to drive a massive gas outflow, at least in this class of systems, though it may have power enough to drag and accelerate clumps to very high velocities.\footnote{See e.g., the recent work by Murthy et al. (2022) where they have detected  a low-luminosity radio galaxy, B2 0258+35, with a massive molecular outflow entirely driven by the radio jet.}

We  quantify, in particular, observed quantities like the mass outflow rate, the filling factors of different gas components, and the duty cycle of the AGN activity, and also compare  our results with previous simulations and several observed classes of active galaxies, including young radio galaxies with molecular and ionized outflows (e.g. \citealt{Morganti2013, Morganti2015}; \citealt{Dasyra&Combes2016A&A...595L...7D}; \citealt{Oosterloo2017...608...A38}; \citealt{Maccagni2018...614...A42}; \citealt{Murthy2019...629...A58}; \citealt{Oosterloo2019A&A...632A..66O}; \citealt{Schulz2018...617...A38,Schulz2021...647...A63}; \citealt{Morganti2021}; \citealt{Murthy2022NatAs...6..488M}), Ultraluminous Infrared Galaxies (ULIGs) (\citealt{Sturm2011}; \citealt{Dasyra2012}; \citealt{Veilleux2017}; \citealt{Cazzoli2017}; \citealt{Pereira-Santaella2018a, Pereira-Santaella2021}; \citealt{Chen2020, Chen2020a}; \citealt{Herrera-Camus2020}; \citealt{Almeida2021}; \citealt{Fluetsch2021}), and Seyfert galaxies (\citealt{Fluetsch2019}; \citealt{Tombesi2012a, Tombesi2015a, Tombesi2017}).

This paper is organized as follows: in section \ref{sec:numerical.model} we  describe our numerical model. Section \ref{sec:results} is devoted to describe and analyze the results of our simulations. In section \ref{sec:discussion}, we discuss and compare our results with previous studies and with observations, and finally in section \ref{sec:Conclusions} we draw our conclusions.
%
%
\section{Numerical Model}\label{sec:numerical.model}
We integrate numerically the magnetohydrodynamic (MHD) equations given by, 
\begin{eqnarray} \frac{\partial\rho}{\partial t}+\nabla\cdot(\rho\textbf{\textit{v}})=\dot{m_S}&&\\ 
\frac{\partial\rho\textbf{\textit{v}}}{\partial t}+\nabla\cdot(\rho\textbf{\textit{v}}\textbf{\textit{v}}-\textbf{\textit{BB}}+\textbf{\textit{p}}^{*})=f_G&&\\ 
\frac{\partial E}{\partial t}+\nabla\cdot\left[(E+p^{*})\textbf{\textit{v}}-\textbf{\textit{B}}(\textbf{\textit{B}}\cdot\textbf{\textit{v}})\right]=E_S&&\\ 
\frac{\partial\textbf{\textit{B}}}{\partial t}-\nabla\times(\textbf{\textit{v}}\times\textbf{\textit{B}})=0&& 
\end{eqnarray}
where $\rho$ is the density, $\textit{v}$ is the velocity, $\textit{B}$ is the magnetic field, $\textit{p}^*$ is a diagonal effective pressure tensor with components $p^*=p+B^2/2$ and $E$ is the total energy density given by,
\begin{equation} E=\frac{p}{\gamma-1}+\frac{1}{2}\rho \textit{v}^2+\frac{\textit{B}^2}{2}, \end{equation}
where $p$ is the thermal pressure and $\gamma$ is the ratio of the specific heats. We assumed an equation of state for an ideal gas, $p=(\gamma -1)e$, where $e$ is the internal energy density and $\gamma=5/3$. The source term $\dot{m_S}$ represents the rate of mass injection by supernovae SNe Ia and SNe II, $f_G$ represents the external gravitational forces due to dark mater, the bulge and the disk of the galaxy, and $E_S= E_{SN Ia}+E_{SNII}+E_{AGN}+E_{DM}+E_{disk}+E_{bulge}-\Lambda (\rho, T)$ represents the sources of energy density coming from SN type Ia and SN type II explosions, the AGN wind, and the external potentials due to the dark matter halo, the disk and the stellar bulge as well as the radiative cooling losses, respectively.

In order to integrate these equations numerically, we used the GODUNOV code \citep{MelioliDalPino2015ApJ...812...90M} \footnote{https://bitbucket.org/amunteam/godunov-code/src/master/}, which  was modified in order to introduce the effects of radiative cooling  over a wide range of temperatures, 50K$<T<10^9$K (\citealt{Raga2000,Melioli2004,Townsend2009, Schure2009}).  We considered an optically thin gas in ionization equilibrium with a metal abundance in the midplane given by $z/z_{\odot}=1.0$ and decreasing linearly with the height of the disk up to a minimum value of $z/z_{\odot}=0.1$, and also included molecular hydrogen $H_2$. For integration in space we used the Harten-Lax-van-Leer C (HLLC) Riemann solver for the HD-simulations, and the Harten-Lax-van-Leer D (HLLD) for the MHD-simulations (see \citealt{ToroEleuterio1999}). A second-order Runge-Kutta scheme (\citealt{Liu1998}) was employed for time integration. The radiative cooling function $\Lambda(\rho, T)$  is calculated implicitly in each time step and implies errors smaller than 10$\%$ for the gas cooling down to a minimum (floor) temperature. In all our simulations we have considered a minimum temperature  $T=50$K (see section \ref{sec:results}).

\subsection{Boundary Conditions}
\label{sec:boundary.conditions}

The computational domain has physical dimensions $1\times 1 \times 1 $ kpc$^3$ in the \textit{x, y,} and \textit{z}-directions for most of the models with a collimated AGN wind, and  $1\times 1 \times 2 $ kpc$^3$ dimensions for the models with a spherical (360°) AGN wind as well as for a few models with 10° opening angle (see more details in section \ref{sec:AGN.setup}). We have run most of the models with $256^3$ resolution (corresponding to a cell size $\sim$3.9 pc). Simulations performed by \cite{MelioliDalPino2015ApJ...812...90M} for a similar domain considering resolutions between $128^3$ and $512^3$ have indicated a good degree of convergence in the results for $256^3$ resolution. For this reason, we have adopted this resolution, i.e.,  a cell size $\sim$3.9 pc,  in the present work for most of the runs, but  have also run a few models with resolution twice as large, with cell size $\sim$1.95 pc. Our simulations have outflow boundaries conditions, which ensure that the fluid reaching the boundaries escapes from the system without coming back. 

\subsection{Initial Setup}\label{sec:initial.setup}

The  initial setup  is also similar to \cite{MelioliDalPino2015ApJ...812...90M}, except for the introduction of the magnetic field into the system. The initial distribution of the gas in the disk is obtained from the solution of the equation of motion in stationary state. The initial mass for the galaxy includes the contribution of a bulge and a stellar disk, and then we set the ISM in equilibrium with the gravitational potential, $\Phi(r,z)$, given by the sum of the dark matter halo, the bulge, and the disk contributions (see eqs. (8)-(10) of \citealt{MelioliDalPino2015ApJ...812...90M}). The galaxy parameters  adopted in this study are presented in Table \ref{galaxy_parameters}.

\begin{figure}
\includegraphics[width=\columnwidth]{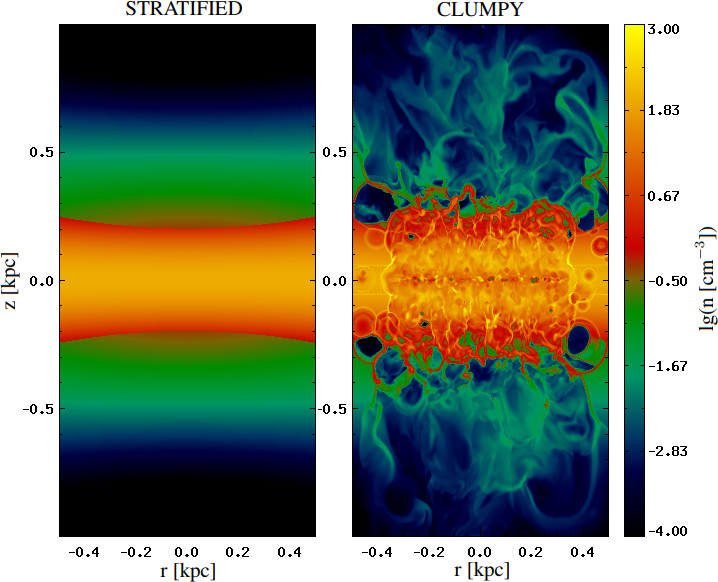}
\caption{2D-cuts in logarithmic scale of the  gas  density distribution showing an edge-on view of the galaxy central region.  Left panel shows an initial smooth distribution which applies to models labeled "smooth" in Table 2. The right panel shows a more evolved ISM (clumpy environment) with a wind that develops after successive SNe explosions, around t = 1.6 Myr. This will serve as initial setup for the models labeled  "clumpy" with low star formation rate in Table 2 (see text for more details). }\label{hmg.inhmg.initial}
\end{figure}

We consider an initially stratified multiphase interstellar gas distribution, with each $i$-phase  defined by a typical temperature $T_{i}$, which is initially in hydrostatic equilibrium in the gravitational potential as described above. The initial gas density distribution for this $smooth$ setup  is shown on the left panel of Figure \ref{hmg.inhmg.initial}. 

In order to account for the presence of magnetic fields in our simulations, we assume the fluid to be initially embedded either in a constant horizontal magnetic field or in a stratified one assuming  initial constant ratio of the thermal to magnetic pressure $\beta= P_{th}/P{m}$. With this prescription, eq. 11 of \cite{MelioliDalPino2015ApJ...812...90M}  is modified as follows:

\begin{equation}
    \rho_i(r,z)=\rho_{i,0}\exp\left[-\frac{\phi(r,z)-e^2\phi(r,0)-(1-e^2)\phi(0,0)}{c_{s,i}^2(1+1/\beta)} \right]
\end{equation}
where $c^2_{s,i}$ is the isothermal sound speed of the i-phase of the gas and $e_i$ quantifies the fraction of rotational support of the ISM (see table \ref{galaxy_parameters}). In our model the i-phase density is replaced by the (i+1)-phase density wherever the (i+1)-phase total (thermal + magnetic) pressure is larger than the (coldest and densest) pressure of the i-phase.

For models where we assumed initially constant magnetic field, we adopted only possible value $B_0 =$ 0.76$\mu$G. In the runs with stratified magnetic fields, we consider three initial values of the thermal to magnetic pressure ratio, $\beta=p/(B^2/8\pi) = \infty$, 300, and 30, corresponding to initial magnetic fields at the plane of the disk $B_0 = 0.0\mu$ G (HD models), 7.6$\mu$G, and 24$\mu$G, respectively (where $B_0$  is the magnetic field strength  at the galactic plane). 

These values are compatible with observed values in nearby Seyfert galaxies and ULIRGs which point to average values of $\sim 1-100 \mu$G, with the latter value occurring only in starburst systems with very large SFRs (\citealt{Thompson2006}, \citealt{Drzazga2011}). They are also comparable with those adopted in previous 3D-MHD studies (\citealt{Rodenbeck2016}, \citealt{Ntormousi2018}).
%
{\setlength{\tabcolsep}{25pt}
\begin{table*}
\caption{Parameters common to all the simulations: \textbf{(1)} Radius of the bulge  (which contains $50 \%$ of the stellar mass of the system);  \textbf{(2)-(4)} Centrifugal rotational support factor for each  of the three initial gas phases;  \textbf{(5)} Mass of the central bulge;  \textbf{(6)} Mass of the central SMBH (related to (5) via M$_{\text{bulge}} \sim 10^3$ M$_{\text{BH}}$;  \textbf{(7)} Mass of the virial. 
}
\begin{tabular}{ccccccc} \hline 
  (1)&(2)&(3)&(4)&(5)&(6)&(7) \\
$r_{b}$ & $e_1$ & $e_2$ & $e_3$ & M$_{\text{bulge}}$ & M$_{\text{BH}}$ &  M$_{\text{vir}}$ \\
1.3kpc &        1   &      0.9  &      0.5 &    $10^{10}$ M$_{\odot}$   &    $10^{7}$ M$_{\odot}$ &   $10^{11}$ M$_{\odot}$  \\ \hline
\end{tabular}
\end{table*}\label{galaxy_parameters}
}


{ \setlength{\tabcolsep}{1.5pt} 
 \begin{table*}
\caption{Simulated models in this work. From left to right: \textbf{(1)} name of the model, \textbf{(2)} initially smooth or clumpy environments; \textbf{(3)} initial thermal to magnetic pressure ratio at the disk mid-plane, $\beta_0 = \beta(z=0; t=0)$ (which is constant all over the system for models with initially stratified B-field); \textbf{(4)} initial magnetic field at the disk mid-plane, $B_0=B(z=0)$ (which is constant all over the system except for models with initially stratified B-field); \textbf{(5)} SNR supernovae rate; \textbf{(6)} SFR, star formation rate; \textbf{(7)} Column density N$_H$; \textbf{(8)}  M$_{\text{core}}$, initial gas mass in  the core region $r<40$pc; \textbf{(9)}  M$_{\text{disk}}$, initial gas mass in the disk region $|z|<200$pc; \textbf{(10)}  AGN injection angle: 0° (very collimated), 10° (conical)  and 360° (spherical); \textbf{(11)} AGN wind luminosity; \textbf{(12)} AGN wind mass injection rate; \textbf{(13)} AGN wind injection velocity; \textbf{(14)} AGN wind active phase injection time; \textbf{(15)} resolution and dimensions of the computational domain.}
  \begin{tabular}{lcccccccccccccc} \hline
\textbf{Model} & ISM & $\beta_0$ & $B_0$ & SNR & SFR & log N$_H$&M$_{\text{core}}$ & M$_{\text{disk}}$ & AGN angle & L$_{\text{AGN}}$ & $\dot{M}_ {\text{AGN}}$ & v$_{\text{AGN}}$ & t$_{\text{Active}}$&res.  \\ 
  &      &      &    ($\mu G$)   &    (yr$^{-1}$)  &   (M$_{\odot}$ yr$^{-1}$)  &   (cm$^{-2}$)  &   (M$_{\odot}$)   &    (M$_{\odot}$)   &     &   (erg s$^{-1}$)  &   (g s$^{-1}$)   &   (cm s$^{-1}$)  &   (kyr)  &   \\
  &       &      &      &     &     &     &   ($10^5$)   &    ($10^8$)    &     &   ($10^{42}$)  &   ($10^{23}$)   &   ($10^9$)  &     &    \\ 
    (1)&(2)&(3)&(4)&(5)&(6)&(7)&(8)&(9)&(10)&(11)&(12)&(13)&(14)&(15) \\ \hline
    \textbf{Low SFR} & & & & & & & & & & & & & &  \\  
    S.HD.1SF.0°.L1                     & smooth  & $\infty$ &  0.0  &   0.01   &  1     &  23    &  6.56  &  2.26  & 0°  & 3.54 & 2.36 & 2.0 & 225 & $256^3$ \\ 
    S.HD.1SF.10°.L1                    & smooth  & $\infty$ &  0.0  &   0.01   &  1     &  23    &  6.56  &  2.26  & 10° & 3.54 & 2.36 & 2.0 & 225 & $256^3$ \\ 
    S.MHD.1SF                          & smooth  & $6\cdot10^3$ &  0.76 &   0.01   &  1     &  23    &  2.35  &  2.21  & - & - & - & - & - & $256^3$ \\
    S.MHD.1SF.0°.L1                    & smooth  & $6\cdot10^3$ &  0.76 &   0.01   &  1     &  23    &  6.56  &  2.26  & 0°  & 3.54 & 2.36 & 2.0 & 225 & $256^3$ \\ 
    S.MHD.1SF.10°.L2            & smooth  & $6\cdot10^3$ &  0.76 &   0.01   &  1     &  23    &  6.56  &  2.26  & 10° & 23.5 & 2.36 & 3.7 & 225 & $256^3$ \\ 
    S.MHD.1SF.10°.L1                   & smooth  & $6\cdot10^3$ &  0.76 &   0.01   &  1     &  23    &  6.56  &  2.26  & 10° & 3.54 & 2.36 & 2.0 & 225 & $256^2$*512 \\ 
    C.MHD.1SF.10°.L2                      & clumpy      &$6\cdot10^3$  &  0.76  &  0.01    &  1     &  23    &  3.35  & 2.28  & 10°    & 23.5 & 2.36 & 3.7 & 225     & $256^2$*512 \\ 
    C.MHD.1SF.360°.L2                     & clumpy      &$6\cdot10^3$  &  0.76  &  0.01    &  1     &  23    &  3.35  & 2.28  & 360°  & 23.5& 30.9 & 3.9 & 75     & $256^2$*512 \\ \hline 
    \textbf{High SFR} & & & & & & & & & & & & & &     \\     
    C.MHD.10SF                         & clumpy      &$1.8\cdot10^4$&  0.76  &  0.1     &  10    &  23.5  &  11.4  & 5.81  & - & - & - & - & - & $256^2$*512 \\  
    C.MHD.10SF.360°.L2                    & clumpy      &$1.8\cdot10^4$&  0.76  &  0.1     &  10    &  23.5  &  11.4  & 5.81  & 360°  & 23.5& 30.9 & 3.9 & 150     & $256^2$*512 \\  
    C.MHD.100SF                        & clumpy      &$1.2\cdot10^4$&  0.76  &  1       &  100   &  24    &  59.0  & 10.1  & - & - & -  & - & - & $256^2$*512 \\ 
    C.MHD.100SF.360°.L2                   & clumpy      &$1.2\cdot10^4$&  0.76  &  1       &  100   &  24    &  59.0  & 10.1  & 360°  & 23.5& 30.9 & 3.9 & 225     & $256^2$*512 \\ 
    C.MHD.100SF.360°.L3 & clumpy      & $1.2\cdot10^4$  &  0.76  &  1      &  100  &  24    & 59.0  &  10.1 & 360°   & 235.0 & 30.9 & 12.3 & 100  & $256^2$*512 \\ 
    C.MHD.1000SF                       & clumpy      & $3\cdot10^4$ &  0.76  &  10      &  1000  &  25    &  424.0 & 50.9  & - & - & - & - & - & $256^2$*512 \\  
    C.MHD.1000SF.360°.L2                  & clumpy      & $3\cdot10^4$ &  0.76  &  10      &  1000  &  25    & 424.0  &  50.9 & 360°   & 23.5 & 30.9 & 3.9 & 375   & $256^2$*512 \\ 
    C.HD.1000SF.360°.L2                   & clumpy      & $\infty$ &  0.0   &  10      &  1000  &  25    & 424.0  &  50.9 & 360°   & 23.5 & 30.9 & 3.9 & 375   & $256^2$*512 \\ 
    C.MHD.1000SF.360°.L3 & clumpy      & $3\cdot10^4$ &  0.76  &  10      &  1000  &  25    & 424.0  &  50.9 & 360°   & 235.0 & 30.9 & 12.3 & 190   & $256^2$*512 \\ 
    C.MHD.1000SF.360°.L2.\textit{high-res} & clumpy      & $3\cdot10^4$ &  0.76  &  10      &  1000  &  25    & 424.0  &  50.9 & 360°   & 23.5 & 30.9 & 3.9 & 375   & $512^2$-1024\\ 
    \textbf{B-field-stratified models} & & & & & & & & & & & & & &\\
    C.MHD$\beta_{300}$.1000SF.360°.L2   & clumpy      & 300 & 7.6 &  10       &  1000   &  25    &  424.0  &  50.9 & 360°  & 23.5& 30.9 & 3.9 & 375  & $256^2$*512\\  
    C.MHD$\beta_{30}$.1000SF.360°.L2  & clumpy      & 30 & 24 &  10      &  1000  &  25    & 424.0  &  50.9 & 360°   & 23.5 & 30.9 & 3.9 & 375   & $256^2$*512\\  \hline
\end{tabular}\label{models}
\end{table*}
}

\subsubsection{SF wind parameters}

Star formation drives turbulence and the development of a clumpy ISM. In order to account for this and also for the possibility of occurrence of star-formation driven outflows in the central regions of the galaxy, we have considered the feedback from massive stars in the late stages of their lifetime when they explode as type II supernovae (SNe II), and also  from low-mass white-dwarfs in binary systems which explode in type Ia SNe, particularly in the bulge.

The rate of SN Ia explosions is proportional to the bulge mass, $\text{SNIa} \sim 0.01(\textsc{M}_{\textsuperscript{bulge}}/10^{10}\text{M}_{\odot})\text{yr}^{-1}$ \citep{Pain1996ApJ...473..356P} which in turn is proportional to the mass of the central supermassive black-hole (SMBH, $\textsc{M}_{\textsuperscript{BH}}$) given  by the relation $\textsc{M}_{\textsuperscript{BH}}/\textsc{M}_{\textsuperscript{bulge}} \sim 10^{-3}$ 
(\citealt{Haring.Rix}), so that we can obtain this rate in terms of the SMBH mass (\citealt{MelioliDalPino2015ApJ...812...90M}): 

\begin{equation}
\text{SNIa} = 10^{-9} \left(\frac{\text{M}_{\textsuperscript{BH}}}{ \text{1M}_{\odot}} \right) \text{yr}^{-1}. 
\end{equation}

To introduce the SNe II, we consider a starburst (SB) in the central region of the host galaxy distributed in the disk within a radius of 300 pc and with a typical half-height (above and below the disk) of 200 pc. Assuming the Salpeter initial mass function (IMF), the number $N$ of stars with mass larger than 8 M$_{\odot}$ (i.e., the SNe II number) is $N \sim 0.01$(M$_b$/M$_{\odot}$), where M$_b$ is the total mass of the stars in the SB. The SNe II activity lasts up to a time t$_b \sim 3 \times 10^7$ yr, which is the lifetime of an 8 M$_{\odot}$ star. A constant SN rate is thus given by $R$ = $N/t_b$, which is in good agreement with more accurate evaluations given by \cite{Leitherer1999ApJS..123....3L}.

We have constrained the initial amount of gas in the disk by the star formation rate using the Kennicutt-Schmidt relation, SFR $\propto n^{\sim 1.4}$  (\citealt{KennicuttJr.1998}) which matches observations for a wide range of active galaxies. Consequently, using the relationships presented above, each model is characterised by a specific column density to which it is associated a given SN rate. We explore here models with SFR between 1 and 1000 M$_\odot$ yr$^{-1}$.

\subsubsection{AGN wind setup}
\label{sec:AGN.setup}
We assume an AGN outflow driven by the SMBH activity in the center of the galaxy. We consider three different geometries for this injection: (i) a collimated outflow with 0$^{\circ}$ opening angle, (ii) a conical outflow with an opening angle of 10$^{\circ}$, and (iii) a spherical outflow injection. In  the cases (i) and (ii), a total luminosity (or mechanical power) $L1=3.5\cdot10^{42}$ erg s$^{-1}$ or $L2=2.35\cdot10^{43}$ erg s$^{-1}$ is injected symmetrically in a volume of $\sim (7.8 pc)^3$, which corresponds to 8 cells in the centre of the computational domain (4 cells above and 4 cells below the mid-plane of the galaxy) characterized by an injection velocity $v_{AGN}$ perpendicular to the plane of the disk and a total rate of injected  matter $\dot{M}_{\text{AGN}}$, wich are given in Table \ref{models}.
In the case (iii), we also consider two possible values for the total injected power  $L2=2.35\cdot10^{43}$ erg s$^{-1}$ or $L3=2.35\cdot10^{44}$ erg s$^{-1}$, introduced in the system as thermal power, which represents a good physical approximation to an adiabatic (energy-driven) propagation of the AGN wind from inner parsec scale \citep{Zubovas2012ApJ...745L..34Z}. In both cases, the wind velocity is given by $v_{AGN}=(2 L/\dot{M}_{\text{AGN}})^{1/2}$.

\subsubsection{Initially smooth and clumpy environments}

The AGN wind is injected in the system considering two different initial setups. In a family of models denominated \emph{Smooth} (S), the wind is injected when the system is in an initially stratified state, in  hydrostatic equilibrium (see left panel of Figure \ref{hmg.inhmg.initial}). In this case, the AGN wind propagates at the same time that the outflow driven by star formation with SN explosions starts to develop. In another family of models denominated \emph{Clumpy} (C), the AGN wind is launched when the system has already developed a clumpy state due to star-formation and the SNe-induced outflow (see right-panel of Figure \ref{hmg.inhmg.initial}).

The initial conditions and parameters of the models analyzed are provided in Table \ref{models}. The model labels use S or C to indicate initially smooth or clumpy environments, respectively; HD represents pure hydrodynamical models, MHD  magneto-hydrodynamical models with an initial constant magnetic field ($B_0 = 0.76 \mu$ G), and MHD$\beta_{number}$ denotes MHD models with an initial constant $\beta=30, 300$; $number$SF stands for the star formation rate (SFR) in units of M${\odot}$ yr$^{-1}$; the subsequent $number$ in the model label specifies the AGN wind opening angle (0°, 10°, or 360°); and the last $number$ indicates the AGN wind luminosity: L1, L2, or L3, corresponding to L${AGN}$= 3.5, 23.5, or 235 $\times 10^{42}$ erg s$^{-1}$, respectively. Finally, the only model presented in the table that was run at higher resolution is labeled as $high-res$.
We have grouped  the models into two samples. One includes the models with star formation rate 1 $M_\odot$ yr$^{-1}$, named $Low$ $SFR$,  most of which have collimated AGN winds, i.e., with 0° or  $10^{\circ}$, characteristic of Seyfert-like systems. This sample has been sub-divided in two categories smooth and clumpy models.
The second big sample  (labeled $High$ $SFR$) includes all models with higher SFR and spherical AGN wind injection, which are typical characteristics of ULIRG-like systems. All the models in this group are initially clumpy.


\section{Results}\label{sec:results}

In this section we discuss the main results or our study, focusing on the feedback of the SB and AGN outflows into the 1 kpc$^3$ central region of the host galaxy over the first 2.0 Myr of evolution. In particular, we will  analyse the influence of the initial conditions in the evolution of the galaxy gas.

\subsection{Convergence of our simulations}
\begin{figure}
\centering
\includegraphics[width=\columnwidth]{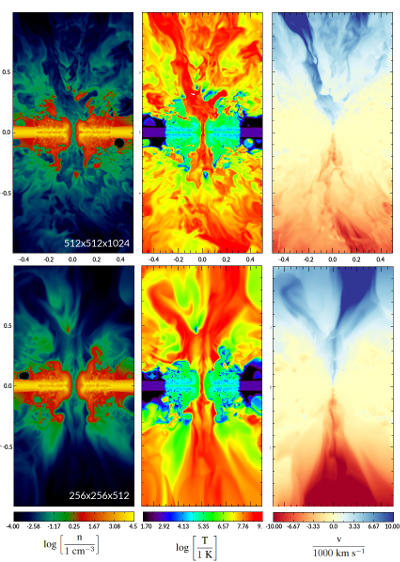}
\caption{\textbf{Numerical resolution convergence.} 2D logarithmic cuts of number density (left), temperature (middle), and velocity (right) for the model C.MHD.1000SF.360°.L2 at t=300 kyr comparing two different grid resolutions: $512\times512\times1024$ (top) and $256\times256\times512$ (bottom). 
} \label{Resolutionx4}
\end{figure}

\begin{figure}
\centering
\includegraphics[width=.7\columnwidth]{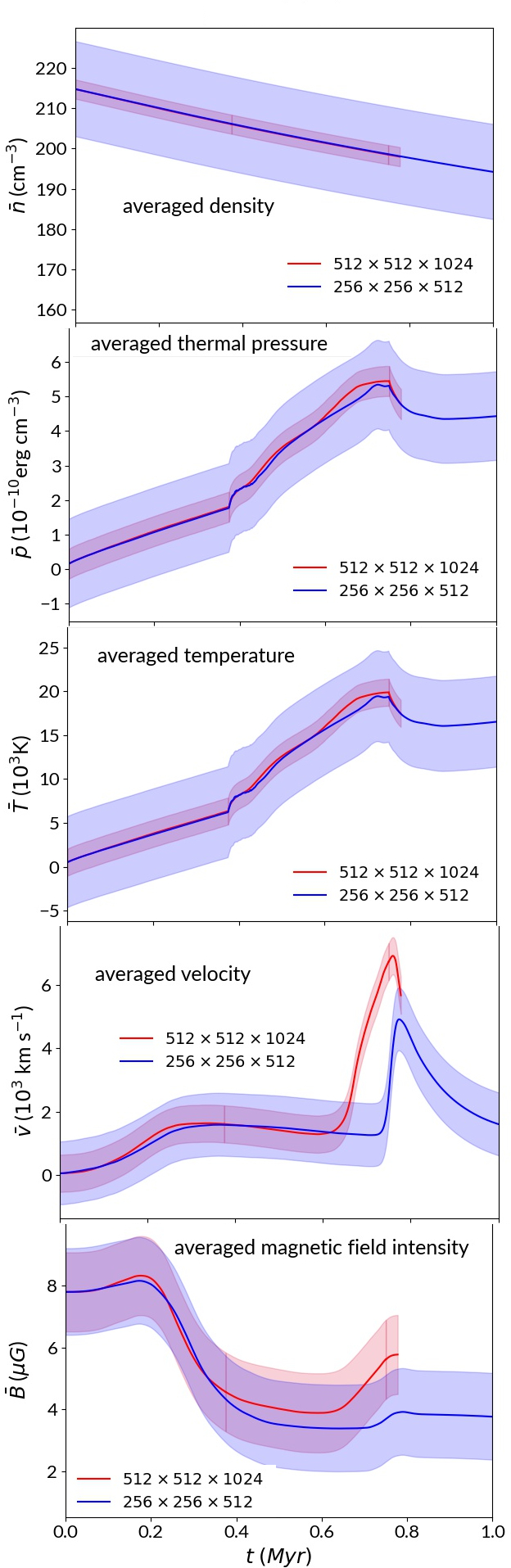}
\caption{\textbf{Numerical resolution convergence.} From top to bottom, volume-averaged density, thermal pressure, temperature, velocity, and magnetic field intensity for model C.MHD.1000SF.360°.L2. The solid red line stands for the higher resolution ($512\times512\times1024$) and the blue for the lower resolution run ($256\times256\times512$). The shaded regions give the variance of each determination.} \label{convvergence.average}\end{figure}

Let us start discussing briefly the convergence of our results. Figure \ref{Resolutionx4} compares the temperature maps for model C.MHD.1000SF.360°.L2 of Table \ref{models} (i.e. with an initial SF rate SF=1000 M$_{\odot}$ yr$^{-1}$, an initial constant horizontal magnetic field with value 0.76 $\mu$G, a spherical (360°) AGN wind and luminosity L2) for two distinct resolutions $512^2\times1024$, corresponding to a cell size $\Delta l = 1.95$ pc in each direction, and $256^2\times512$, corresponding to a cell size $\Delta l = 3.9$ pc in each direction. 
Figure \ref{convvergence.average} compares the evolution of volume averaged quantities for these two models. 

We note that the general features of the evolving system are comparable in both resolutions, as well as the volume averaged quantities. We will discuss all these properties in more detail in the next sections, but the results in these figures reveal good convergence of the results already for a resolution of $256^2\times512$, 
and for this reason, most of the models considered in this work were run with this resolution.
The higher resolution maps in Figure \ref{Resolutionx4}  evidence the natural presence of more features, particularly in the temperature and velocity maps in the halo regions, but the macro-dynamical properties are essentially preserved, as we see also in Figure \ref{convvergence.average}. We only note a slightly delay in the formation and the strength of the peak of the average velocity due to the AGN-wind injection which encounters more resistance to break into the smoother (less porous) lower-resolution environment, but this will not affect the overall evolution of the system.

\subsection{Effect of ISM initial conditions on AGN-wind propagation}\label{sec:initial.conditions.ON.AGN-wind}

\subsubsection{Smooth vs. Clumpy  environments}\label{sec:homog.vs.inhomg}

\begin{figure}
\centering
\includegraphics[width=.8\columnwidth]{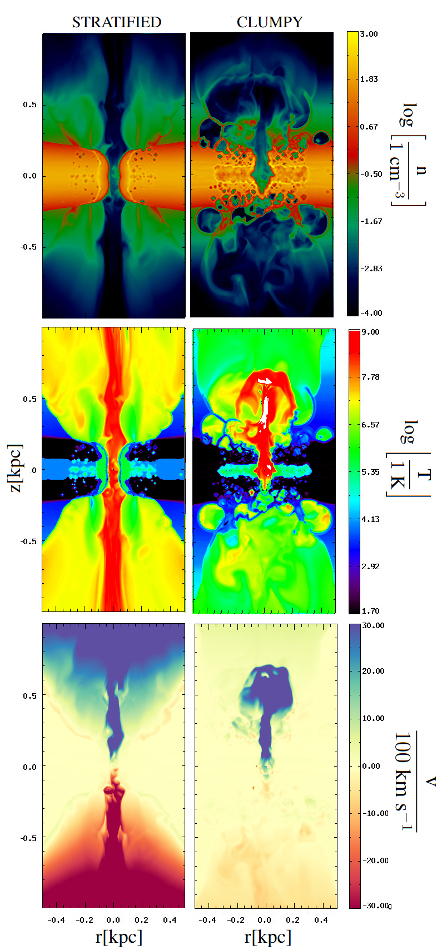}
\caption{\textbf{Initially smooth versus clumpy environments}. This figures shows 2D-logarithmic density (top), temperature (middle), and velocity (bottom) maps of the nuclear region of a system after the evolution of the AGN-wind injected in an initially smooth environment in hydrostatic equilibrium (model S.MHD.1SF.10$^{\circ}$.L2 in table \ref{models}; left panels), and in a clumpy ISM driven by star-formation (model C.MHD.1SF.10$^{\circ}$.L2; right panels). The time depicted is $t=300$ kyr. Both models have an initial horizontal magnetic field $B=0.76$ $\mu$G, a SF rate of SFR=1 M$_{\odot}$ yr$^{-1}$, and luminostiy L2.}\label{strat-clumpy}\end{figure}

In this section we show how the porosity of the ISM induced by star formation affects the feeding and the dynamics of the AGN-wind as it evolves. Porosity is one of the most important characteristics of this study where the multi-phase ISM is not induced by an imposed initial density distribution function, but is instead, a natural consequence of the stellar feedback.

The initial gas distribution determines the evolution of the system. The main differences between an initially smooth and a star-formation induced clumpy environment are shown in Figure \ref{strat-clumpy}. Here we compare density, temperature, and velocity distributions, at $t\simeq 300$ kyr, of two systems where a cone-shaped AGN wind with 10° opening angle is injected in an initially smooth system in hydrostatic equilibrium and in an evolved clumpy ISM, both characterized by an initial horizontal magnetic field $B=0.76 \mu G$ and a SF rate=1 M$_{\odot}$ yr$^{-1}$.
We clearly see that in the smooth environment, the AGN outflow expands faster, easily disrupting the disk, reaching  the halo, and forming a low density, hot regular bipolar-shaped wind.  On the other hand, in the case of the clumpy environment, an asymmetric, wiggling outflow develops on both hemispheres, propagating at slower velocity through the irregular structures that form in the thicker disk due to SF activity. In this case, the AGN outflow barely reaches the halo. The SF driven galactic wind is in turn much broader, but much slower.
Also striking is the fact that when the disk gas is smooth and not dominated by stellar feedback (left panel), the AGN wind expands through the galaxy disk without modifying it significantly and no clouds (or cold gas) are  transported to above the disk. On the contrary, when the AGN wind impinges directly into the inhomogeneous, clumpy ISM, formed as a consequence of the stellar feedback (right panel), the AGN wind has more difficulties in expanding in the disk due to  the action of external ISM pressure gradients. The gas evolution is thus dominated by the stellar activity, and high density, low temperature structures, like filaments and clouds, are more easily pushed into the halo, both by the SB and by the AGN wind.
Star formation rate is therefore, a fundamental parameter in these systems because it determines the efficiency of the stellar feedback and the consequent multi-phase structure of the gas in the disk. A galaxy with negligible stellar activity will produce no significant amount of clouds or filaments in the disk and, consequently will hardly  produce, for instance, molecular gas outflows. For this reason, in the analysis that follows we  consider, in general, reference models from table \ref{models} with a galactic disk setup in which the stellar feedback has already created a clumpy environment and a multi-phase ISM.
\subsubsection{Duty cycle of the AGN activity}
\label{sec.duty.cycle}
Both, the stellar feedback and  the AGN wind inject into the ISM of the disk a large amount of energy which accelerates a fraction of the gas to the halo of the galaxy, and part of it returns, but  to the external regions of the disk. After a given time, which depends on the initial conditions of the system and  the amount of energy injected, the core region of the galaxy will lose almost completely its gas mass, forcing an interruption of the active phase of the AGN. What is evident from our simulations is that the active phase of the AGN cannot be continuous, but must follow alternating cycles of outflow activity and inactivity.

Indeed, as the AGN wind sweeps through, the interstellar (IS) gas is completely expelled from the core. However, during the inactive phase of the AGN wind (when it is turned off), the IS matter is able to recommence its inflow towards the central region. This sets the stage for a new active phase of the AGN. This process  referred to as the "duty cycle" mechanism
(\citealt{Schawinski2015}, \citealt{Zubovas2016}, \citealt{Morganti2017b, Morganti2017c}, \citealt{Zubovas2018}), determines a sort of AGN-galaxy co-evolution, and though it is not yet fully understood, it is widely accepted that the gas content plays a crucial role in triggering both AGN and star formation activity.

Figure \ref{512res.CLUMPY.50K.spher} illustrates this duty cycle for the higher  resolution model  C.MHD.1000SF.360$^{\circ}$.L2.\textit{high-res}. The figure depicts maps of density, temperature,  and velocity for three different snapshots of the AGN activity. At $t= 300$ kyr (active phase), the AGN has disrupted the disk, pushing IS gas from the core, and reached the halo forming a spectacular hot bubble that would be probably observable in X-rays. After a $\sim 350$ kyr active period, the AGN-wind is turned-off because the inflow of ISM fuel to the nuclear region has been interrupted by the AGN outflow itself. The panel at $t=450$ kyr shows the remnant phase of the wind, which is followed by a fully inactive phase (starting after $t\sim 525$ kyr and ending around $t\sim 1125$ kyr) where there is no more trace of the AGN wind-remnant, only a broader outflow in the halo which is mainly driven by the SN bursts. At this stage the inflow of fuel to the core from the active star formation of the ISM is resuming and a new AGN wind phase should start.

\begin{figure}
\includegraphics[width=\columnwidth]{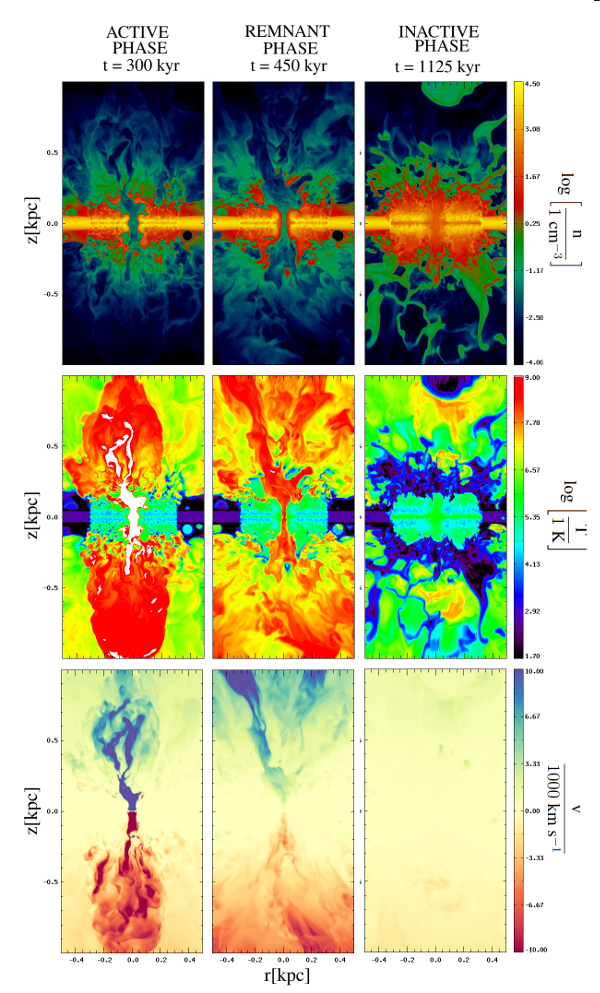}
\caption{\textbf{Duty cycle of AGN activity}. The figure shows
2D-logarithmic maps of the nuclear region of a magnetized system with a
spherical AGN-wind injected in an initially clumpy environment  with SFR=1000 M$_{\odot}$ yr$^{-1}$ (corresponding to the $512\times512\times1024$ resolution model C.MHD.1000SF.360$^{\circ}$.L2.\textit{high-res}).  From top to bottom: density, $n$ ($cm^{-3}$), temperature \textsc{T}  (in K units) and vertical velocity  v$_z/$(1000 km s$^{-1}$). Left panels show  the \textit{active phase} of the AGN wind  at $t=300$kyr; center panels show the \textit{remnant phase}  at $t=450$ kyr, and right panels  show the \textit{inactive phase} of the AGN wind  at $t=1125$ kyr.}
\end{figure}\label{512res.CLUMPY.50K.spher}

Now, let us examine the evolution of the gas mass within the central region of the galaxy, in order to determine a more accurate estimate of the duration of the active phase of the AGN in our models.
\begin{figure}
\includegraphics[width=\columnwidth]{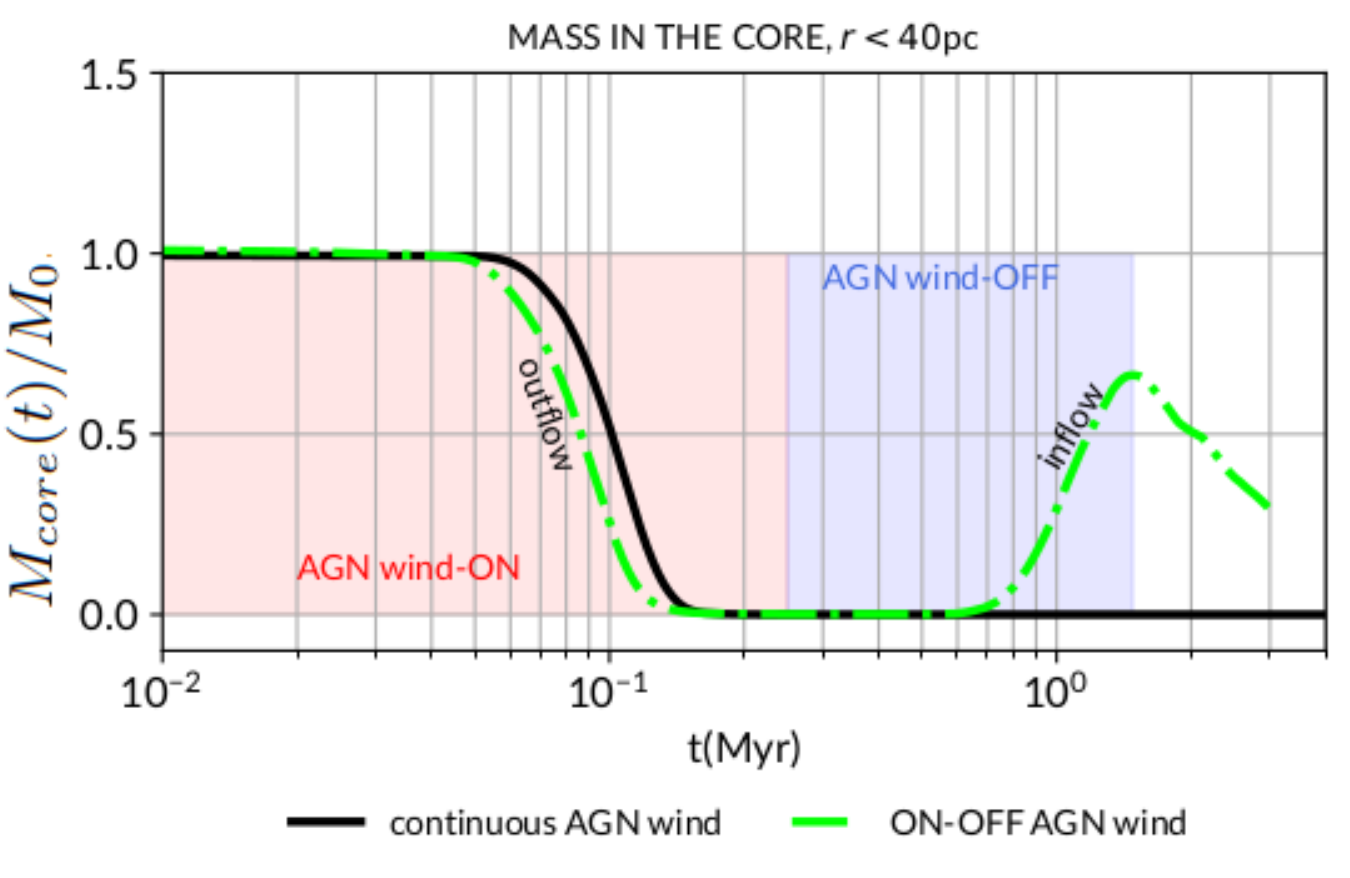}
\caption{\textbf{Characterization of the AGN activity cycle for an initially smooth environment.} This figure shows the fraction of gas in the core region as a function of time  $M_{core}(t)/M_0$.  $M_0$ is the initial gas mass in the core region, $M_{core}(t=0) = 6.57\cdot 10^5 M_{\odot}$. Black line represents a model with  continuous injection of the AGN-wind. The outflow in this case removes totally the  gas from the nuclear region in $\sim 10^5$yr and does not allow for  further refuelling of the core region. The green line represents a wind that is interrupted after this period (AGN-off in the plot). Later on, a gas replenishment is observed in the nuclear region that reaches a maximum value lower than the initial mass in the core, but could trigger a new active phase of the AGN wind. The initial conditions  correspond to model S.MHD.SF.10°.L1 (see Table \ref{models}).} \label{mass.core.cont.interm} 
\end{figure}
As an example, figure \ref{mass.core.cont.interm} shows the time evolution of the average mass fraction $(M(t)_{core}/M_0)$ in the central region of the galaxy for the smooth model S.MHD.SF.10°.L1, where $M_0$ is the initial gas mass in the core region. 
The model depicted by the green line illustrates an interrupted AGN wind after a duration of $\sim 2 \cdot 10^5$ years, whereas the black line represents a continuous injection of the AGN wind. In the case of continuous AGN wind, the gas within the central region is entirely expelled after approximately $2 \cdot 10^5$ years. The high-pressure exerted by the AGN wind prevents the inward flow of interstellar gas into the central region, thus interrupting the fuel supply to the supermassive black hole (SMBH) and subsequently halting the AGN outflow. This suggests that a continuous AGN wind injection beyond approximately $2 \cdot 10^5$ years is no longer realistic. Consequently, we conducted the same model with a constrained active phase duration of the AGN wind at around $2 \cdot 10^5$ years (green line).

In the absence of AGN wind energy injection, the gas stellar feedback generates an inflow of gas that gradually fills the central region of the galaxy. Eventually, the replenished gas reaches a maximum value, which may trigger a new phase of AGN wind activity. This pattern exemplifies the duty cycle of the AGN wind, which, depending on the initial conditions of each model, may exhibit slight variations in duration. However, our simulations consistently indicate a range of a few $10^5$ years for both smooth and clumpy environments (see table \ref{models}).

\begin{figure}\includegraphics[width=\columnwidth]{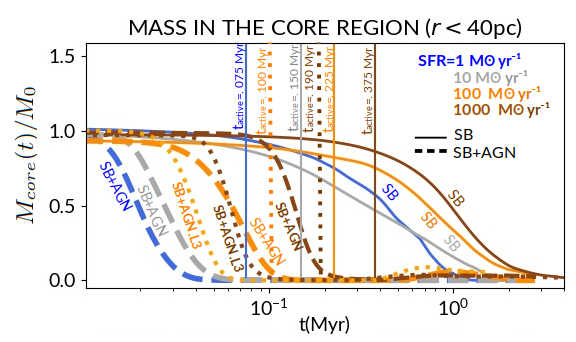}\caption{\textbf{Characterization of the AGN activity cycle for initially clumpy environments with different SFR and different AGN wind luminosities}. The gas mass is depicted as a function of time in the core region, $r<40pc$, for initially clumpy models. Solid lines stand for models hosting a SB wind only, and dashed lines for both, SB+AGN winds.  Blue, grey, orange and brown lines stand for models with AGN wind luminositiy L2 and star formation rates: SFR=1, 10, 100, and 1000 M$_{\odot}$yr$^{-1}$, respectively. The vertical lines indicate the time at which the AGN-wind is turned-off in each of these models: 75 kyr, 150 kyr, 225 kyr and 375 kyr, respectively. The dotted orange and dotted brown curves show the same, but  for the models with AGN wind luminosity L3 and SFR = 100 and 1000 M$_{\odot}$yr$^{-1}$, respectively. For these models, the dotted vertical lines indicate an AGN-wind turn-off at 100 kyr and 190 kyr, respectively. For each model the gas mass is normalized to the initial mass in the core, $m_{core}=M(t)/M_0$ (see table \ref{models}). All the models depicted have initial spherical AGN wind, and initial constant magnetic field 0.76 $\mu$G (see table \ref{models}).}\label{mass.core.ULIRG-like}\end{figure}

We note that these values are consistent with several observations (see, e.g., \citealt{Schawinski2015} and references therein). 
In view of the results above, we focus in this work on models in which the injection of the AGN wind is turned off when the fuel provided by the IS gas inflow is interrupted by the wind itself. This allows us to follow at least one duty cycle of the AGN feedback into the galaxy. An illustration of this is shown in Figure \ref{mass.core.ULIRG-like} for several initially clumpy models of table \ref{models} with different SFR. The core mass evolution is followed until the end of an AGN wind cycle, i.e., right before a new cycle of gas inflow to the nuclear region and a new activity of AGN wind should resume. In the figure we also highlight (in the vertical blue, grey, yellow and brown lines) the characteristic time ($t_{\rm {active}}$) at which the AGN wind is turned-off due to the gas emptying in the central region.  These values are also listed in Table \ref{models}. 

It is evident that the duration of AGN-wind activity exhibits a clear correlation with the star formation rate (SFR), reflecting the amount of gas present in the disk or the column density, denoted as $N_H$. Additionally, we observe that the active phase of the AGN decreases, as expected, with the power of the AGN wind. The impact of AGN wind power on the active phase is relatively larger compared to the variation in SFR. For instance, when considering a SFR of 1000 M$_{\odot}$yr$^{-1}$, we note that the increase of the AGN wind luminosity by a factor of 10 (from L2 to L3) leads to a  49\% decrease in the AGN activity time, reducing it from 375 kyr to 190 kyr. In contrast, for the same luminosity L2, decreasing the SFR by a factor of 10 (from 1000 to 100 M$_{\odot}$yr$^{-1}$) results in a reduction of the activity time from 375 kyr to 225 kyr, amounting to a 40\% decrease.
Figure \ref{mass.core.ULIRG-like} also shows another important result. The dashed-dot lines highlight the gas evolution in the central region for models hosting stellar feedback only, with no AGN wind.
Although stellar feedback plays a key role in the creation of a multi-phase medium in the disk of the galaxy (as shown in sec. \ref{sec:homog.vs.inhomg}) and is, therefore, essential in promoting the formation of dense,  filamentary structures (see discussion in the next section), it is evident that its action alone results to be much less efficient in removing gas from the central region of the galaxy, taking much longer time for that, of at least $\sim$3 Myr. For this reason, the duty cycle in our models is clearly  more correlated  with the active phase of the AGN wind.
\subsubsection{AGN wind opening angle}\label{sec:Opening.Angle}
We are also interested in understanding the role that the AGN wind plays on the evolution of the host galaxy. For this reason, we have considered three different setups for the AGN wind: $i$) a collimated outflow perpendicular to the disk of the galaxy with 0° opening angle; $ii$) a collimated outflow perpendicular to the disk of the galaxy with  10° opening angle; $iii$) and a spherical wind  corresponding to an opening angle of 360° (which resembles observed Ultra Fast Outflows - UFOs).

The  simulations with 0° and 10° opening angle revealed no substantial differences, a  result which is also sustained in higher resolution simulation (performed for model C.MHD.1SF.10°.L1) with  resolutions 256$^2*$512 and 512$^2*$1024,  compared with its 0° counterpart at resolution 256$^3$).

\begin{figure*}
\includegraphics[width=\textwidth]{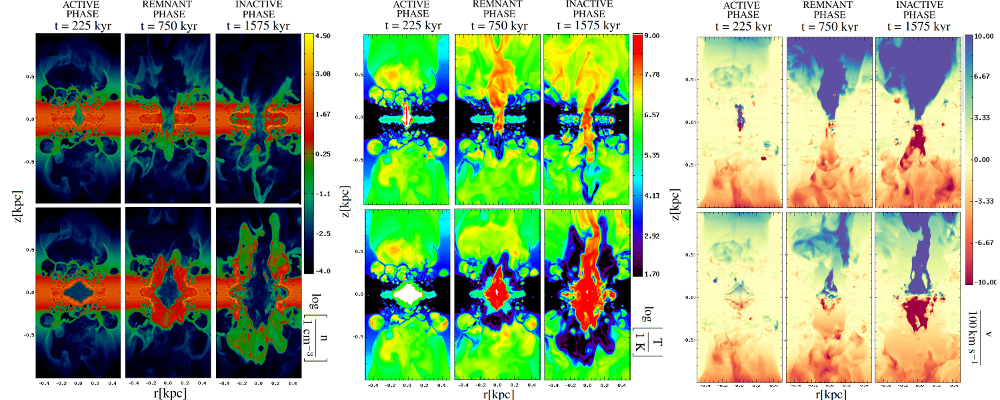}
\caption{\textbf{Influence of AGN jet opening angle.} 2D edge-on logarithmic maps of the central slice comparing the AGN wind opening angle effects on the production of outflows. From left to right, each set of panels contains the density, temperature and vertical velocity, respectively, of the models C.MHD.1SF.10°.L2 (top panels)  and C.MHD.1SF.360°.L2 (bottom panels)  at evolutionary times from left to right: $t=$ 225, 750 and 1575 kyr.
} 
\label{TOTAL.JET.vs.SPHER}
\end{figure*}

However, it is essential to conduct a more meticulous analysis of the differences between collimated and spherical wind models. To achieve this, we examine two specific models from table  \ref{models}: C.MHD.SF1.10°.L2 and C.MHD.SF1.360°.L2, both with identical initial conditions, except for the wind opening angle, which is 10° and 360°, respectively.
Figure \ref{TOTAL.JET.vs.SPHER} shows the evolution of the density, temperature and velocity, for these two models. During the first 775 kyr, the energy deposited by the AGN wind heats the ISM, which in turn starts to expand doing work on the circumnuclear gas. For the 360° AGN wind
the working surface is larger and thus it drags more material during the expansion. At the end of the activity phase (at $t=0.225$ Myr), the remnant of the 10° AGN wind propagates rapidly through the disk and is deposited in the halo region, mixing with the hot gas. The same occurs for the spherical AGN wind, but at smaller speed, and it mixes with the hotter gas in the halo region only after about 1 million year. 
The reason why this happens is that, unlike the 10° AGN wind, for which the energy is deposited mostly in the halo region, part of the energy of the spherical AGN wind is used to push the dense gas in all directions of the disk plane.

\subsubsection{Star Formation Rate}\label{sec:effect.SFR.on.AGN}

In this section, we explore the evolution of the gas of the galaxy as a function of the SFR, 
in order to understand how it influences the formation of high density structures, such as  clouds and filaments, in the outflow.
The relation between the amount of gas available in the disk to form stars and the star formation rate is well known:  SFR $\propto n^{\sim 1.4}$  (\citealt{KennicuttJr.1998}). This is equivalent to have an increase in the SFR with the column density of the gas ($N_H$), i.e., the integrated density along the disk.
As remarked, we have performed numerical simulations considering  SFR ranging from  1 M$_{\odot}$yr$^{-1}$ to 1000 M$_{\odot}$yr$^{-1}$
(Table \ref{models}). In order to ensure the
correspondence with the Kennicutt–Schmidt law \citep{KennicuttJr.1998}, the central gas density in these models varies between 10$^{2}$ $-$ 10$^{4}$ cm$^{-3}$, respectively.

\begin{figure}
\includegraphics[width=\columnwidth]{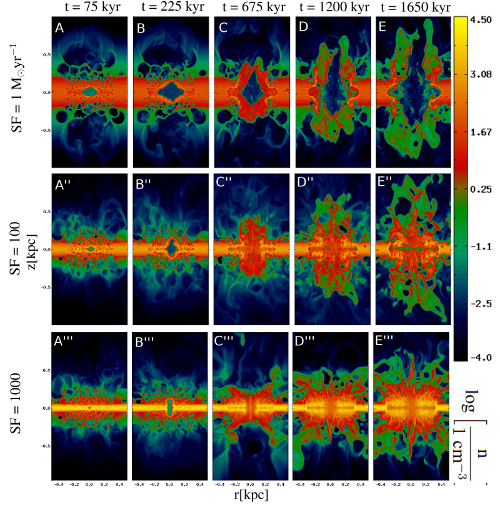}
\caption{\textbf{Dependence with the SFR.} Depicted are 2D edge-on logarithmic maps of the central slice of the number density $n(cm^{-3})$ for the models C.MHD.1SF.360°.L2 (top), C.MHD.100SF.360°.L2 (middle) and C.MHD.1000SF.360°.L2 (bottom), with SFR=1 M$_{\odot}$ yr$^{-1}$, 100 M$_{\odot}$ yr$^{-1}$, and 1000 M$_{\odot}$ yr$^{-1}$, respectively. From left to right: $t=$75kyr,225kyr,675kyr,1200kyr and 1650kyr.}
\label{DN.SFR.ON.AGN}
\end{figure}

\begin{figure}
\centering
\includegraphics[width=.6\columnwidth]{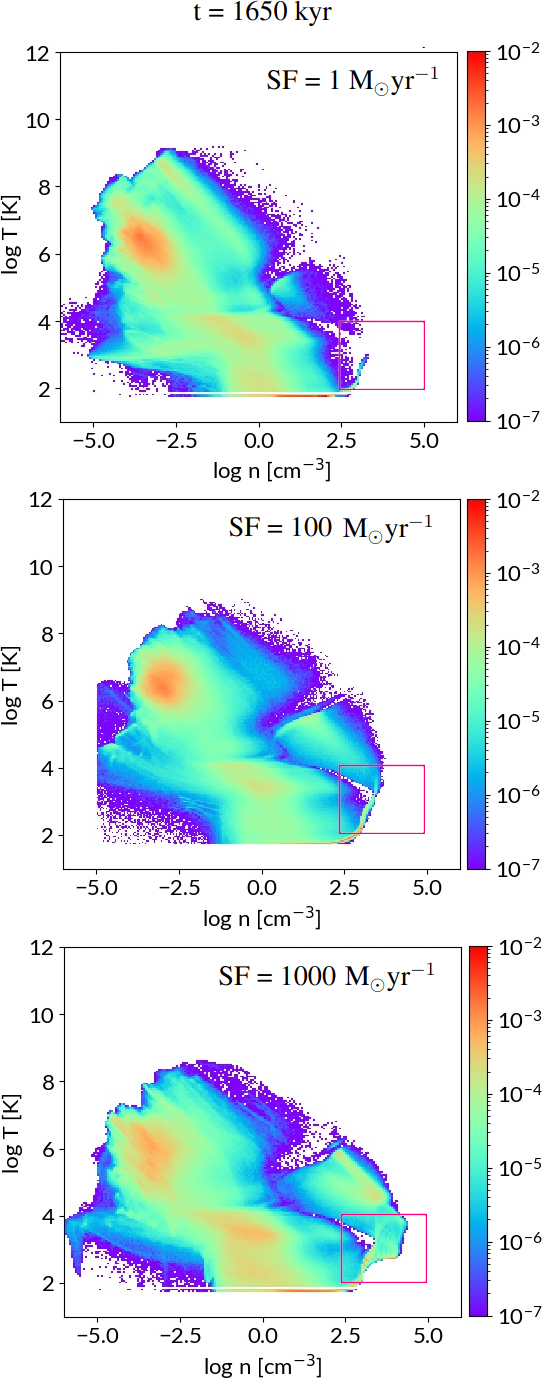}
\caption{\textbf{Dependence with the SFR.} Shown are 2D-histograms of density versus temperature, $n$ vs $T$, for the same models as in Figure \ref{DN.SFR.ON.AGN}, C.MHD.1SF.360°.L2 (top), C.MHD.100SF.360°.L2 (middle), and C.MHD.1000SF.360°.L2 (bottom), for $t = 1650$ kyr, comparing the effect of different SFR on the AGN wind propagation and feeding. The pink box in the bottom right of each panel indicates the densest and coldest gas in the systems.}
\label{hst.nT.SF.ON.AGN}
\end{figure}

In Figure \ref{DN.SFR.ON.AGN} we show the gas density distribution evolution for three models with spherical AGN wind ejection with different initial SFR=1, 100, and 1000 M$_{\odot}$yr$^{-1}$. 
We note three main differences. First of all, the higher density of the disk in the model with higher SFR naturally slows down the expansion of the AGN wind. This effect is also enhanced by the higher pressure of the ISM induced by the higher number of SN explosions, directly connected with the SFR. The second difference is  the larger number of dense structures formed in the disk in this case. This characterizes how the porosity (or "clumpiness") of the disk, increases  with the SFR and will be studied in detail in section \ref{sec:volume.filling.factor}, where we evaluate the filling factors of the multi-phase components of the gas, but in Figure \ref{hst.nT.SF.ON.AGN} we can already distinguish the larger number of colder, denser structures in the model with larger SRF. Lastly, the third distinction pertains to the evacuation process of the central region of the galaxy. The larger density and SFR facilitate the inward movement of interstellar gas towards the nuclear region, consequently delaying the emptying process and promoting a more effective mixing of the ISM with the AGN wind. This phenomenon also accounts for the longer active phase of the AGN wind observed in models with higher SFR (see section \ref{sec.duty.cycle} and Figure \ref{mass.core.ULIRG-like}; see also table \ref{models}).

\begin{figure}
\includegraphics[width=\columnwidth]{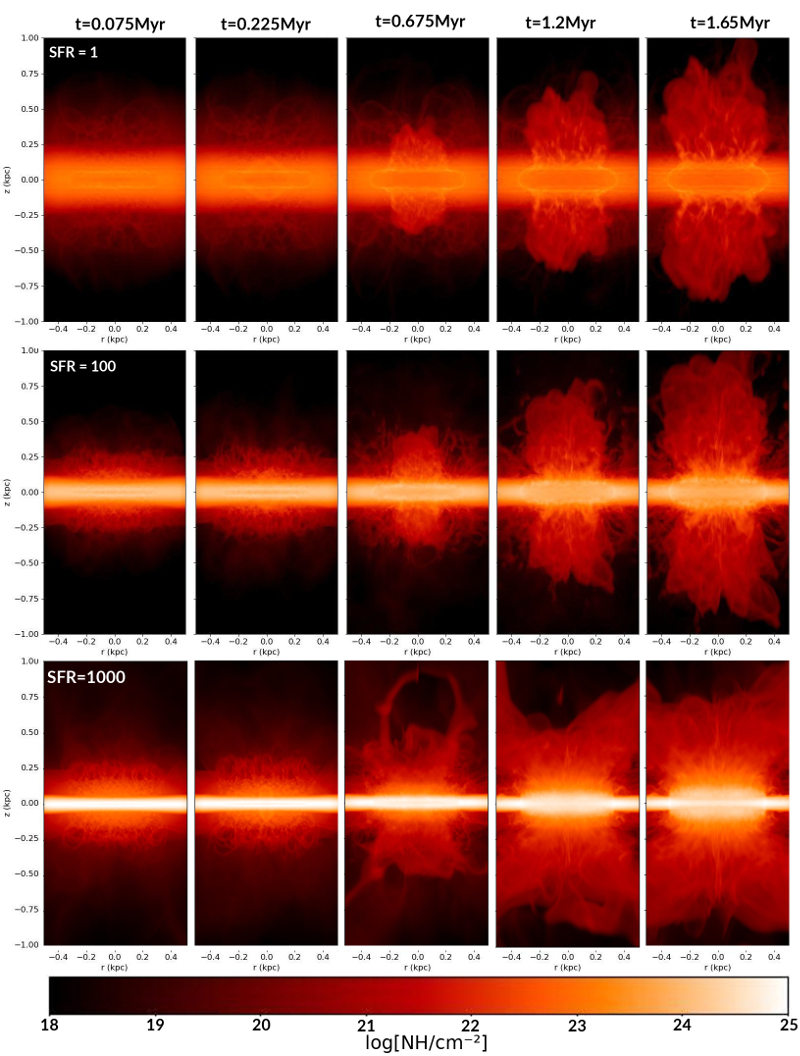}
\caption{\textbf{Dependence with the SFR.} The figure shows 2D edge-on logarithmic maps of the column density $N_H (cm^{-2})$ for the models of  Figure \ref{DN.SFR.ON.AGN} integrated along a line of sight parallel to the disk, at $X$ direction. 
} \label{colden.SFR1-100}
\end{figure}

Figure \ref{colden.SFR1-100} shows the column density maps integrated along the normal direction to the disk for the same models of Figure \ref{DN.SFR.ON.AGN}. The column density, which is a quantity that can be directly  observed, obviously blurs the structuring features we see in the density maps. Nonetheless, we can still appreciate the intrinsic differences due to the increase of the SFR, particularly in the disk region.

SFR is, therefore, one of the most sensitive parameters of our models, and for this reason in section \ref{sec:volume.filling.factor} we will focus mainly on this parameter for the study of the evolution of the gas of the galaxy and in the comparison with observations in section \ref{sec:observations}.

\subsubsection{HD vs MHD models}\label{sec:HD.vs.MHD}

\begin{figure*}
\includegraphics[width=\textwidth]{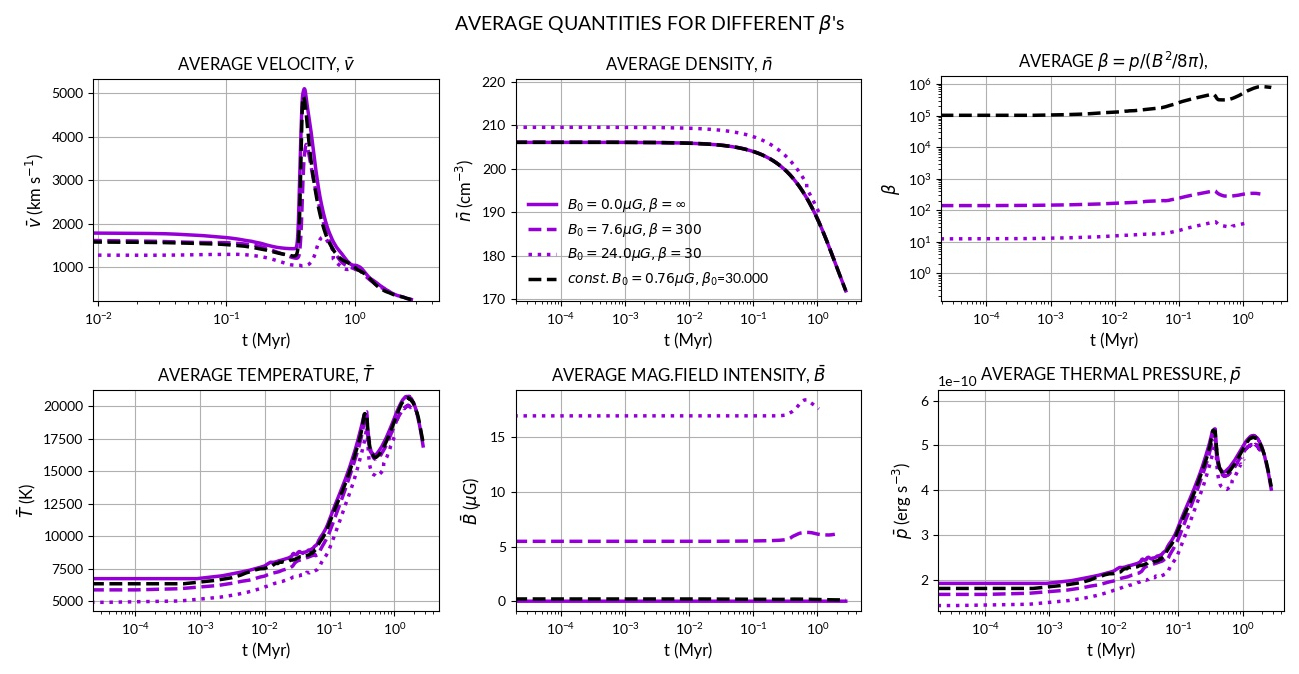}
\caption{\textbf{HD versus MHD}. Volume averaged quantities for the model C.MHD.1000SF.360°, which has initial constant magnetic field in the entire domain, $B_0=0.76 \mu$G, corresponding to a value of $\beta$ at z=0 given by $\beta= 30.000$ (dashed black line). It is  compared with the HD model ($\beta= \infty$; continuous line), and  the models with initial stratified magnetic fields with  $\beta=$ 300 and 30, corresponding  to a magnetic field at z=0  $B_0=7.6$ $\mu$G, and $B_0=24$ $ \mu$G, respectively (dashed and dotted purple lines).
Top panels, from left to right: volume averaged velocity ($\bar{v}$), number density ($\bar{n}$), and thermal to magnetic pressure ratio ($\bar{\beta}$)
Bottom panels: volume averaged temperature ($\bar{T}$), magnetic field intensity ($\bar{B}$), and thermal pressure ($\bar{p}$). 
}
\label{v.n.T.HD.vs.MHD}
\end{figure*}

\begin{figure}\includegraphics[width=\columnwidth]{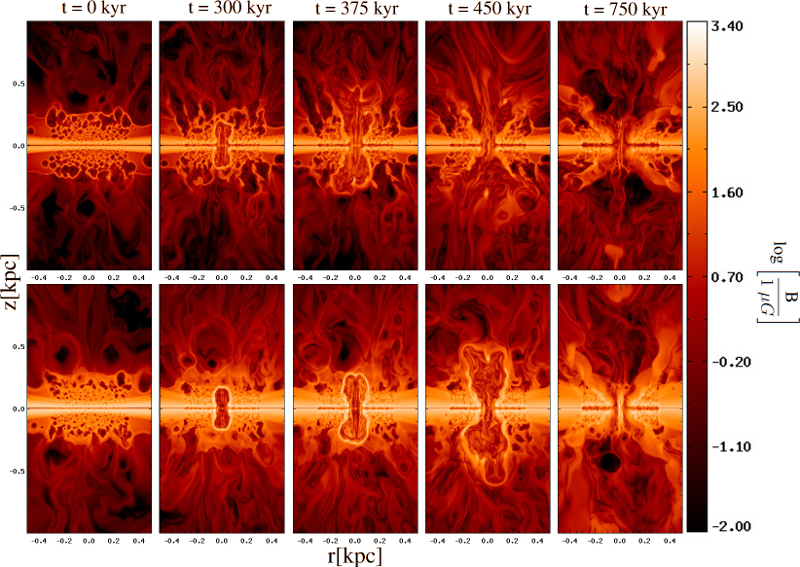}\caption{\textbf{HD versus MHD.} 2D edge-on logarithmic maps of the central slice of the magnetic field intensity ($B$) for the models C.MHD$\beta_{300}$.1000SF.360°.L2 with $\beta=$300 (top), and C.MHD$\beta_{300}$.1000SF.360°.L2 with $\beta=$30 (bottom) shown in five different instants (from left to right): $t=0,300,375,450,750$ Myr.} \label{B.HD.vs.MHD}\end{figure}
         
\begin{figure}
\includegraphics[width=\columnwidth]{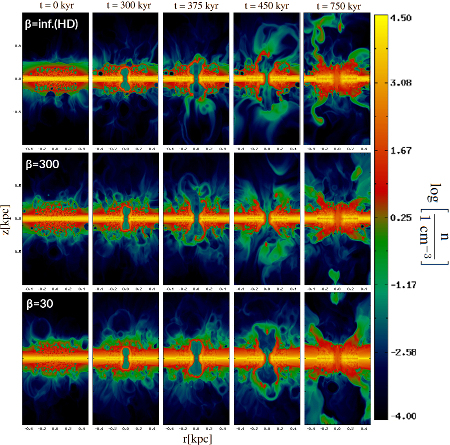}
\caption{\textbf{HD versus MHD}. 2D edge-on logarithmic maps of the central slice of the density ($n$) for the models 
HD.1000SF.360°.L2 ($\beta=\infty$, top),  C.MHD$\beta_{300}$.1000SF.360°.L2  ($\beta=$300, middle) and  C.MHD$\beta_{30}$.1000SF.360°.L2 ($\beta =$30, bottom), 
shown in five different times from left to right: $t=0, 300, 375, 450, 750$ Myr. 
}
\label{DN.HD.vs.MHD}
\end{figure}

\begin{figure}
\includegraphics[width=\columnwidth]{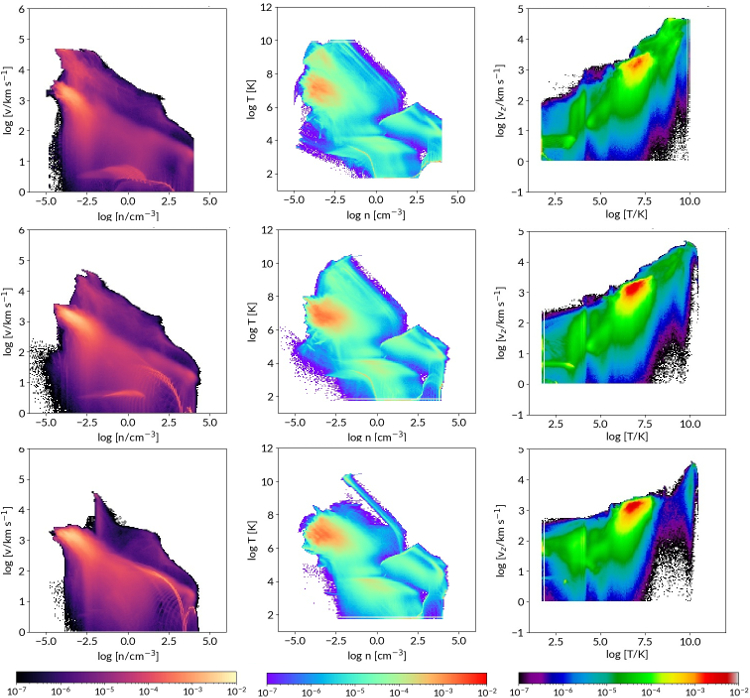}
\caption{\textbf{HD vs MHD.} Two-dimensional histograms of the number density vs. vertical velocity (left panels), number density vs. temperature (middle panels), and temperature vs. vertical velocity (right panels) of the gas distribution for the models 
HD.1000SF.360°.L2 ($\beta=\infty$, top),  C.MHD$\beta_{300}$.1000SF.360°.L2  ($\beta=$300, middle) and  C.MHD$\beta_{30}$.1000SF.360°.L2 ($\beta =$30, bottom), at a time $t$=0.375 Myr.
}  
\label{hst.nv.HD.vs.MHD}
\end{figure}

In this section, we compare HD and MHD models to better understand the role that the presence of  magnetic fields can play on the evolution of the system, specially in star-formation induced clumpy environments. 

In Figure \ref{v.n.T.HD.vs.MHD} we compare the evolution of  volume averaged  quantities of the system, such as velocity ($\bar{v}$), density  ($\bar{n}$), temperature  ($\bar{T}$), magnetic field intensity ($\bar{B}$), and thermal pressure ($\bar{p}$). 

For the set of initial horizontal stratified magnetic fields considered here, with $\beta=$  30, 300, and $\infty$ (the latter corresponding to the hydrodynamical system), or an initial constant field with $B_0 =$ 0.76 $\mu$G, the comparison of the average quantities indicates no substantial differences between the models. In fact, the system with an initial constant $B_0 =$ 0.76 $\mu$G, which is the value adopted in most of the models of this work (see Table \ref{models})  has essentially the same average density, temperature, pressure and vertical velocity of the HD counterpart. The same applies to the stratified model with $\beta=$ 300, which has a maximum initial magnetic field 10 times larger at the disk. Some minor differences on the average velocity, density, temperature and pressure become more noticeable in the model with $\beta=$ 30, which corresponds to a maximum initial magnetic field $B_0=24 \mu$G at the disk. The average total velocity decreases from 5000 km $s^{-1}$ in the $B_0 =$ 0.76 $\mu$G and HD models, to less than 2000 km $s^{-1}$ in the $\beta=$ 30 model, caused by the deceleration of the AGN wind due to the impact into the magnetized gas which absorbs part of the kinetic energy of the terminal shock. The same applies to the average temperature and thermal pressure which decrease with increasing magnetic field (decreasing $\beta$), due to the absorption of part of the internal energy density in the compression of the magnetic field. 
Additionally, we have found  that when we intensify the initial magnetic field  to $B_0=76 \mu$G in the disk ($\beta=$ 3), the aforementioned distinctions become significantly more pronounced, especially after the impact of the AGN wind, reflecting tangible dynamical influence of the magnetic fields in the system's evolution. While we have not presented the results of these stronger magnetized models in this study, we intend to delve into them in a forthcoming work.

Figures \ref{B.HD.vs.MHD} to \ref{hst.nv.HD.vs.MHD} confirm the trends above revealed by the average quantities. Figures \ref{B.HD.vs.MHD} and \ref{DN.HD.vs.MHD} compare central 2D slices of the magnetic field intensity and density distribution, respectively, for the models with initial stratified magnetic fields $\beta=$ 300 and 30, and the HD model ($\beta=\infty$). Figure \ref{hst.nv.HD.vs.MHD}, compares two-dimensional histograms of the number density vs. vertical velocity, number density vs. temperature, and temperature vs. vertical velocity of the entire gas distribution for the same models. Clearly, no substantial differences are seen among these models, except for the occasional presence of small filaments and clumps with larger-than-average magnetic field strength  accumulated there by compression, specially for the model with $\beta=30$. In this case, there is a slightly larger quantity of high velocity (up to few $\sim 100$ km/s), 
colder and denser structures (see Figure \ref{hst.nv.HD.vs.MHD}).
In summary, the results above indicate that magnetic fields with average values up to a few tens $\mu$G, as observed in most of the sources investigated here, have no major effect on the evolution or  on the formation of high-velocity large  dense, cold structures in these systems.
\subsection{Gas mass outflow}
As discussed previously, the expulsion of gas from the galactic disk towards the halo is driven by both stellar feedback (the SB-driven wind) and the AGN wind. This evolutionary progression is highly influenced by the initial conditions of the system (refer to Section \ref{sec:initial.conditions.ON.AGN-wind}) and the radiative cooling of the gas. Conversely, as emphasized in Section \ref{sec:intro}, observations reveal the existence of dense and cold (molecular) gas structures with high velocities outside the disks of numerous active galaxies, indicating that the AGN wind predominantly contributes to the presence of such features at elevated altitudes. Consequently, our focus lies in quantifying the mass loss rate within our models to gain a deeper understanding of which combination of initial conditions and physical processes can effectively describe and substantiate these observational outcomes.

\begin{figure}
\includegraphics[width=\columnwidth]{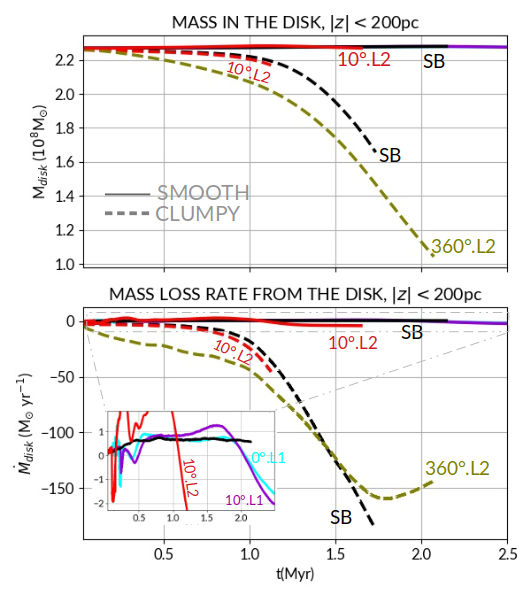}
\caption{\textbf{Gas outflow dependence on initially smooth versus clumpy environment.} The figure shows the mass as function of time for the disk region ($|z|<200$)  (top) and the corresponding mass loss rate (bottom) for  smooth  (continuous line) and clumpy (dashed line) models with 
SFR = 1 M$_{\odot}$yr$^{-1}$, named  Low SFR  in table \ref{models}. Black, green, red, purple, and  cyan,  lines stand for models with: no AGN wind (SB); AGN wind with 360° and luminosity L2; AGN wind with 10° and luminosity L2; AGN wind with 10° and  luminosity L1; and AGN wind with 0° and luminosity L1, respectively. We note that the curves of all  smooth models, which have   AGN wind 0° or 10° practically superpose, even for different wind luminosities. The inset highlights the  very small differences among these models.
}\label{disk.loss}
\end{figure}

In Figure \ref{disk.loss} we show the mass loss rate from the disk region ($|z| < 200$ pc) for the  models labelled as low SFR  in table \ref{models}). 

First and foremost, these plots provide compelling evidence for a bimodal evolution of mass outflow, contingent upon whether the galactic disk exhibits an initial smooth or clumpy gas distribution. It is clearly evident that the gas mass outflow is considerably smaller in the smooth models. Conversely, no significant differences are observed between models with AGN wind opening angles of 0° or 10°, as well as their counterparts without AGN wind (SB curves). However, as expected (refer to Section \ref{sec:Opening.Angle}), a higher mass loss rate is observed in the case of a 360° AGN wind.

\begin{figure}
\centering
\includegraphics[width=\columnwidth]{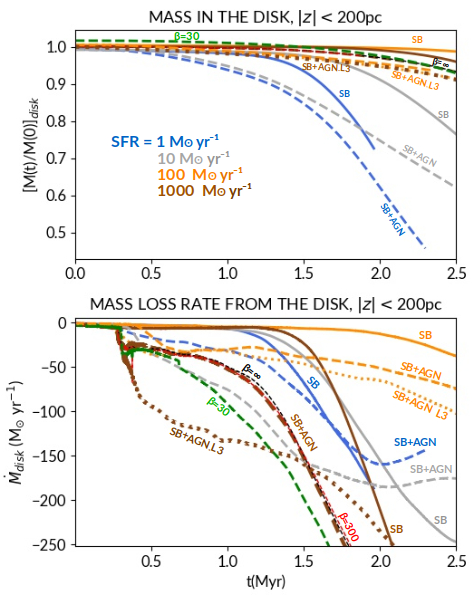}
\caption{\textbf{Gas outflow dependence with SFR}. This figure shows the mass loss (top) and mass loss rate (bottom) from the disk $|z|<200pc$ region for the models in table \ref{models}  with spherical AGN wind injection (or SB wind only) differing by the star formation rate: SFR=1, 10, 100 and 1000 M$_{\odot}$yr$^{-1}$ (blue, grey, orange and brown lines respectively). Continuous lines stand for models hosting SB winds only, and dashed lines for models hosting SB+AG winds. All these models have initial constant magnetic field $B_0 = 0.76 \mu$G and AGN wind luminosity L2 = 2.35$\times 10^{43}$ erg s$^{-1}$. The black, red and green lines stand for the counterpart models to the one  with SFR=1000 M$_{\odot}$yr$^{-1}$ (brown line),  which have initial $\beta = \infty$, 300 or 30, respectively). The models with  $\beta = \infty$ and 300  superpose entirely with the brown line model. Also depicted are the models with SFR=100 and 1000 M$_{\odot}$yr$^{-1}$, and AGN wind luminosity L3= 2.35$\times10^{44}$erg s$^{-1}$ (dotted orange and dotted brown lines, respectively).} \label{disk.loss_sfr}
\end{figure}

Models with stellar feedback + AGN wind characterised by both larger SFR (and thus larger gas column density) and larger opening angle, present the most influential effects on the gas outflow and structure formation. This is shown in Figure \ref{disk.loss_sfr} for the set of models labeled High SFR in table \ref{models}, which are also compared with their counterpart with SFR = 1 M$_{\odot}$yr$^{-1}$. For these models, the total gas mass lost by the central region of the disk (within R=200 pc) varies between 10\% (for  SFR=100-1000 M$_{\odot}$ yr$^{-1}$) and more than 40$\%$ for SFR$\leq$  10 M$_{\odot}$ yr$^{-1}$, of the initial total mass of that region.  The smaller percentage for larger SFR is due to the comparatively larger total amount of gas in the system for increasing SFR. A more complex, nonlinear  behaviour is denoted in the corresponding mass loss rates shown in the bottom diagram. Models with same AGN wind luminosity, but different initial SFR reveal no clear trend, particularly the model  with SFR = 100 M$_{\odot}$ yr$^{-1}$, which has a comparatively smaller rate than its counterpart models. This can be attributed to the fact that for small SFR, there are less SN explosions and structure formation to prevent the outflow expansion to the halo,  while for a large enough SFR, as in the case of SFR=100 M$_{\odot}$ yr$^{-1}$, these obstacles become larger causing the decrease of the mass loss rate, but if one further increases the SFR (to SFR=1000 M$_{\odot}$ yr$^{-1}$), then the expansion of the structures due to much larger number of SN explosions will instead $help$ the outflow propagation thus increasing again the mass loss rate (as seen in the bottom panel), evidencing a clear cooperation of both the SN-driven and the AGN winds. This is what we observe from an inspection of the maps of outflow evolution in these models.
Moreover, the mass loss rate is generally smaller in  models  where only the SB wind is present, compared to their counterparts where the AGN wind is also considered.
The increase of the AGN wind power also affects the mass loss rate from the disk. This can be observed by comparing the two models in Figure \ref{disk.loss_sfr} with SFR=100 M$_{\odot}$ yr$^{-1}$ and AGN wind powers L2 (dashed orange curve) and L3 (dotted orange curve). Notably,  the model with larger power has a mass loss rate twice as large. 

The same applies to the models in the Figure \ref{disk.loss_sfr} with SFR=1000 M$_{\odot}$ yr$^{-1}$. Overall, the models indicate mass loss rates from the disk between 50 and 250  M$_{\odot}$ yr$^{-1}$.

We also notice that for the magnetic field strengths investigated here, there is no significant influence on the mass outflow.

The results above are consistent with those of the previous sections, i.e., they confirm that the  models  which are able to transport a large amount of gas to outside  the disk and allow for the formation of dense structures in the halo are those wherein there is a combined action of AGN and stellar feedback, being sensitive to both the SFR,  and the power and opening angle of the AGN wind, though the dependence on the SFR reveals a more complex trend, that we will try to  better disentangle in the next section.

\subsection{Filling factor}\label{sec:volume.filling.factor}

The analysis above of the gas mass outflow is not enough  to understand the evolution of the gas above the disk, and more specifically, its multi-phase composition and velocity. Therefore, in order to trace the formation of structures in the outflow, we compute the volume filling factor $f_V$ given by the ratio of the  volume $V'$ occupied by a given set of $N'$ structures with a density above a given threshold, to the total volume $V$ of the system ($V'\leq V$). The total volume is given by the box of our computational domain and considering that the volume of the cell is $v_{cell}$, we have

\begin{equation}
f_V=\frac{V'}{V} = \frac{N'v_{cell}}{N_{total}v_{cell}}=\frac{N'}{N_{total}}
\end{equation}

\begin{figure}
\centering
\includegraphics[width=.8\columnwidth]{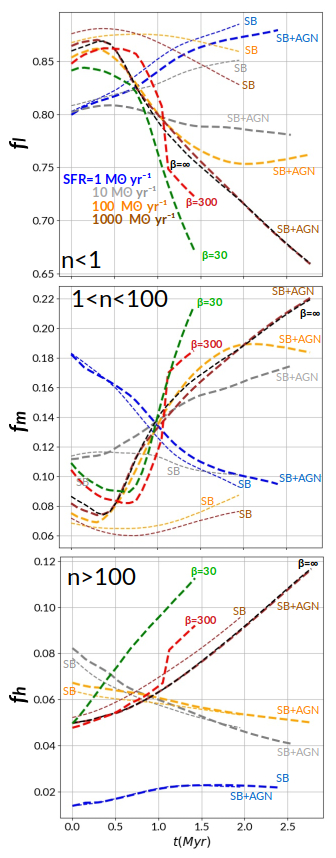}
\caption{\textbf{Volume Filling factor $f_V$ as function of time for systems with different SFR.} The gas is separated in three different ranges giving $f_V=f_l$ for $n<1$ $cm^{-3}$ (top panel), $f_V=f_m$ for 1$cm^{-3}$$<n<100$ $cm^{-3}$ (middle panel), and  $f_V=f_h$ for $n>100$ $cm^{-3}$ (bottom panel). Systems with SFR = 1, 10, 100 and 1000  $M_{\odot}yr^{-1}$  are represented by blue, grey, orange and brown lines, respectively.  Thin lines represent models with stellar wind feedback only (SB wind) and thick lines the models with both SB+AGN winds (see more details in the text). All the models with AGN wind have spherical injection and luminosity L2=$2.35 \times 10^{43}$ erg s$^{-1}$.}
\label{fv.all.in}
\end{figure}

For this study, we have considered three different gas number density ($n$) ranges, namely: (\textbf{\textit{low}}) $f_l=f_V(n<1$ $cm^{-3})$; (\textbf{\textit{medium}})  $f_m=f_V( 1cm^{-3}<n<100$ $cm^{-3})$ and (\textbf{\textit{high}}) $f_h=f_V(n>100$ $cm^{-3})$.
The volume filling factor for the models with spherical AGN wind  of table \ref{models}  which differ mainly in their SFR, are depicted in  Figure \ref{fv.all.in}.

The multi-phase composition of the gas influences the evolution of the outflow and the appearance of dense clouds outside the disk. In previous studies \citep[e.g.][]{Wagner2011, Cielo2017, Asahina2017a, Mukherjee2018a, Mukherjee2018c, Tanner_2022}, 
the distribution of the multi-phase medium was chosen as a free parameter in the initial conditions, whereas in this study this distribution is a direct consequence of stellar feedback. As shown in Figure \ref{fv.all.in}, this distribution evolves with time and changes completely depending on the initial conditions of the system as well as the interactions between stellar and AGN wind. For models with SFR higher than 1, the AGN wind reduces the fraction of low density gas, $n<1$ $cm^{-3}$ (Fig. \ref{fv.all.in}, top ) and together with the continuous action of the stellar feedback, it increases the fraction of medium, $1<n<100$ $cm^{-3}$ (Figure \ref{fv.all.in}, middle) and high density gas, $n>100$ $cm^{-3}$ (Figure \ref{fv.all.in} bottom).

It is interesting to compare our results in Figure \ref{fv.all.in} with those of  \cite{Wagner2011}. In their models, they imposed two fixed values for the initial filling factors, amely 0.13 and 0.42, for  clumps with densities in the range 30 - 1000 cm$^{-3}$, distributed at heights of the order of 500 kpc above the disk. In our models, where the filling factor is naturally built by SF, we obtain initial values  that are at most $\sim$ 0.12, for the same range of density structures, distributed up to a $\sim$ 200 kpc height, at the initial states of  the clumpy models. As time evolves, this quantity decreases to values of the order of 0.01-0.02.   

While  Figure \ref{disk.loss_sfr} lacks information on how the amount of high altitude  dense gas in the galaxy relates to the SFR,  Figure \ref{fv.all.in} provides  more detailed information.
We note that the concentration of dense material will be much more significant for  SFR = 100 and  1000 $M_\odot yr^{-1}$, than for  SFR =  1 or 10 $M_\odot yr^{-1}$, as clearly indicated in the bottom diagram of the Figure \ref{fv.all.in}. In fact, after about 2.5 Myr, the filling factor of the gas with number density $\geq$ 100 cm$^{-3}$ above the disk reaches values between 11\% and 2\% for SRF=1000 and SFR=1 $M_\odot yr^{-1}$, respectively. The intensity of the stellar feedback is therefore, one of the main ingredients to load  the galactic outflow with molecular gas.
We also note that the $\beta=$30 model improves slightly the feeling factor of the high density component by at most a factor $\sim$ 1.7 with regard to the $\beta=$300 counterpart. 

\begin{figure}
\includegraphics[width=\columnwidth]{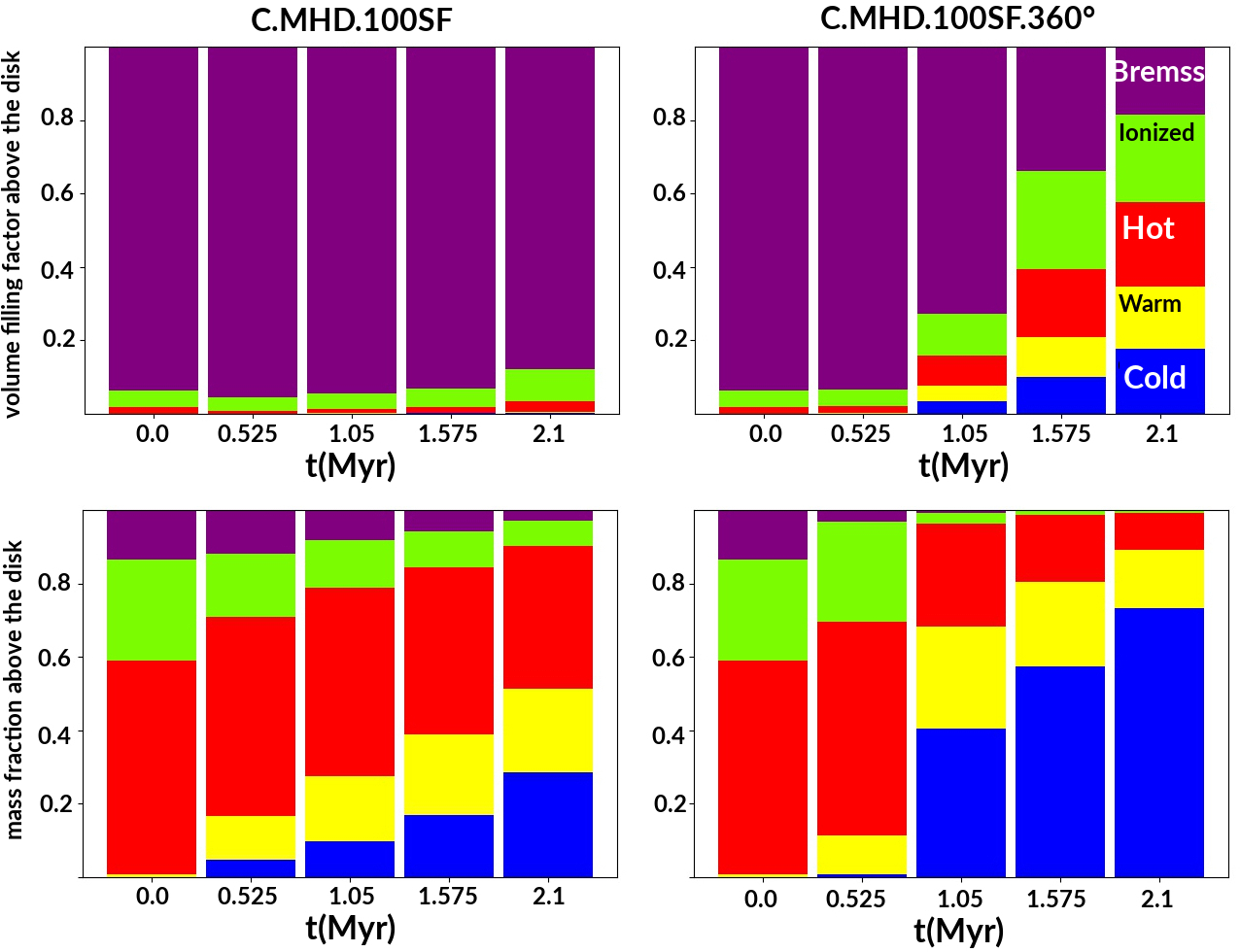}
\caption{\textbf{Volume filling factor} (top panels) and the associated mass fraction  (bottom panels) in each temperature range above the disk on both sides ($|z|> 200$pc) for the models with:  SB wind only  (C.MHD.100SF, left) and with SB+AGN wind (C.MHD.100SF.360$^\circ$.L2, right) shown in five different times:  t=0.0 Myr, t=0.525 Myr, t=1.05 Myr, t=1.575 Myr and t=2.1 Myr. The model with AGN wind removes a bigger fraction of cold  and warm gas (right panels), specially after 2.1 Myr of evolution, when the AGN wind remnant is no longer  visible. This suggests that, at least at the scales studied here ($<$ 1 kpc), the past  AGN wind activity is a determining factor in the production of cold outflows, that are the most massive component of the gas being ejected. 
}
\label{append.ff.mf.NH24}
\end{figure}

\begin{figure}
\includegraphics[width=\columnwidth]{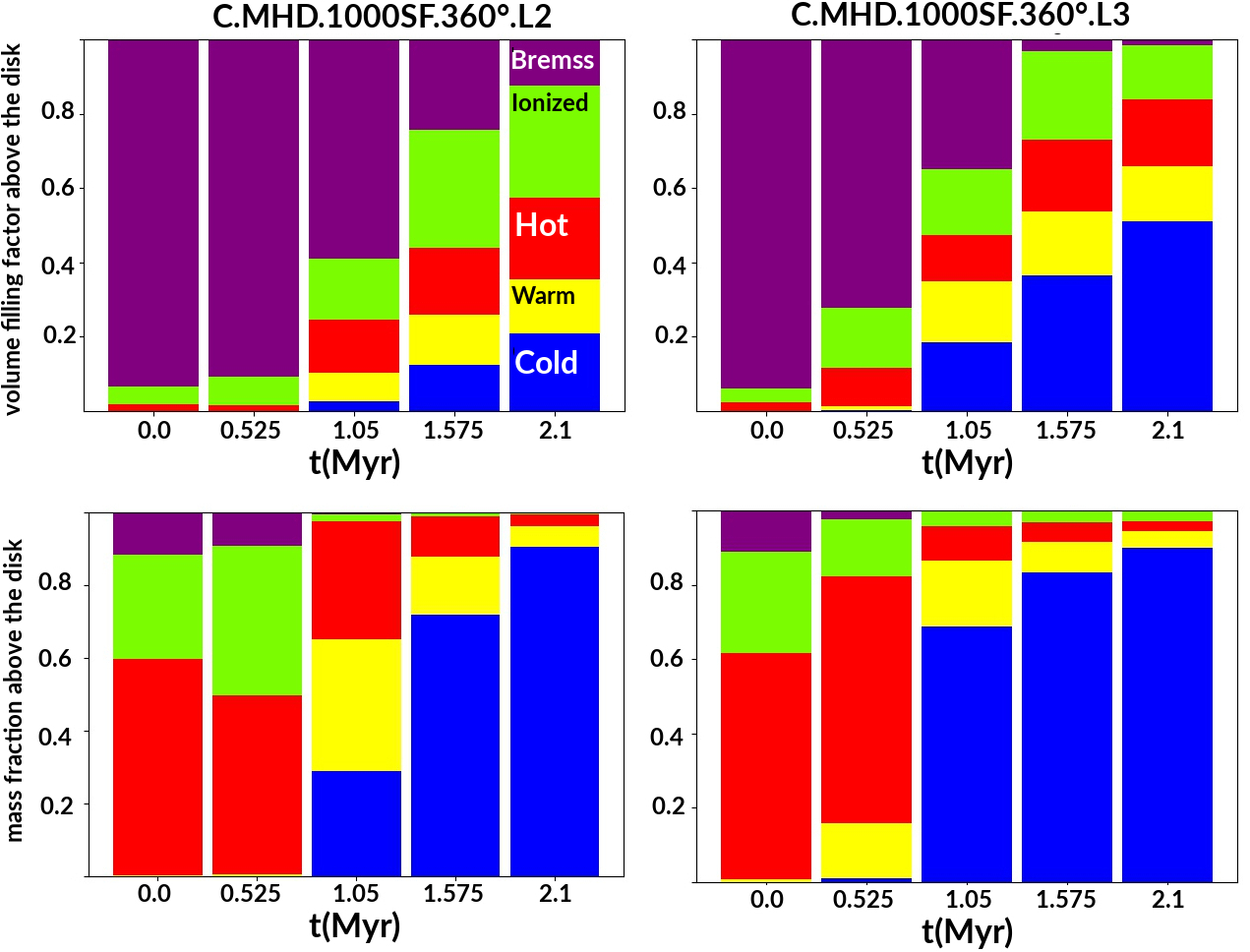}
\caption{\textbf{Volume filling factor} (top panels) and the associated mass fraction in each temperature range (bottom panels) in each temperature range above the disk on both sides  ($|z|> 200$pc) for  models differing only in the AGN wind luminosity: C.MHD.1000SF.360$^\circ$.L2 and C.MHD.1000SF.360$^\circ$.L3 (left and right panels, respectively), which differ only by the power injected by the AGN wind (of $10^{43}$ and $10^{44}$ erg s$^{-1}$, respectively). It is   shown for five different times:  t=0.0 Myr, t=0.525 Myr, t=1.05 Myr, t=1.575 Myr and t=2.1 Myr.  
At $t\sim $ 1.05 Myr, the AGN wind with higher luminosity removes more mass, but after 2.1 Myr, the amount removed (by the SF+AGN wind) is almost the same in both models. 
} 
\label{fT.bars.1000SB.vs.AGN}
\end{figure}

The study of the volume filling factor of the gas is also particularly important when we analyse the temperature distribution of the gas above the disk. To do this, we have classified the gas in the following ranges: cold: (50-200) K, warm: (200-1500) K, hot: (1500, 10$^4$) K, ionized: ($10^4-10^6$) K, bremsstrahlung: ($10^6-10^9$) K and res: (residual) $>10^9$ K. 
In Figures \ref{append.ff.mf.NH24} and \ref{fT.bars.1000SB.vs.AGN}, we explore the mass fraction distribution  above the disk region ($|z|>200$ pc) for the two models that, according to Figure \ref{fv.all.in} are able to transport cold dense (molecular) gas to outside the disk more efficiently, namely, C.MHD.100SF.360$^\circ$ and C.MHD.1000SF.360$^\circ$. 

More specifically, in Figure \ref{append.ff.mf.NH24}, we compare the multiphase gas distribution both, in the absence and presence of the AGN wind (left and right panels, respectively), calculated for the model with SFR=100 $M_\odot yr^{-1}$. The figure shows not only  the volume filling factor (top),  but also the mass filling factor (bottom panels). At $t=0$ Myr, the gas distribution above the disk is the same in both cases and the most massive component is the hot gas (1.500 K-10.000 K) followed by the ionized component, although their volume filling factor is  $\sim 5\%$. 
During the system's evolution, it becomes evident that the SB wind alone, as well as the combined influence of the SB and AGN wind, effectively generate and carry a substantial amount of cold gas beyond the disk's boundaries. Nevertheless, these two winds fulfill distinct roles in the gas evolution. The SB wind primarily contributes to filament formation, whereas the AGN wind predominantly facilitates the transportation of denser structures located above the disk. Consequently, in the absence of the AGN wind in the model, the proportion of cold gas transported above the disk after 2 million years amounts to approximately 30\%. Conversely, when the AGN wind is included in the model, this percentage increases to around 70\%.

Similarly, Figure \ref{fT.bars.1000SB.vs.AGN} depicts the multiphase gas distribution for the model with  SFR=1000 $M_\odot$ yr$^{-1}$, but considering two different injection luminosities for the AGN wind. Clearly, the model with 10 times larger luminosity is able to remove a grater amount of cold gas, approximately 2.7 times larger, corresponding to a total mass of $\sim$ $1.75\cdot 10^7$ M$_{\odot}$. 
 
Moreover, we have seen in Section \ref{sec:HD.vs.MHD} that the AGN wind is able to accelerate cold gas to high velocities. Cold structures with a few 100 km/s are identified for AGN wind luminosities L2 (see Figure \ref{hst.nv.HD.vs.MHD}) and should increase further with increasing  luminosity.
On the other hand, when comparing the model C.MHD.1000SF.360.L2 in Figure \ref{fT.bars.1000SB.vs.AGN}
with its counterpart in Figure \ref{append.ff.mf.NH24} differing only by the SFR=100 $M_\odot$ yr$^{-1}$, we observe that the cold gas above the disk after 2 Myr increases from 70\%  to about $90\%$ from the lower to the larger SFR model.

\begin{figure}
\includegraphics[width=\columnwidth]{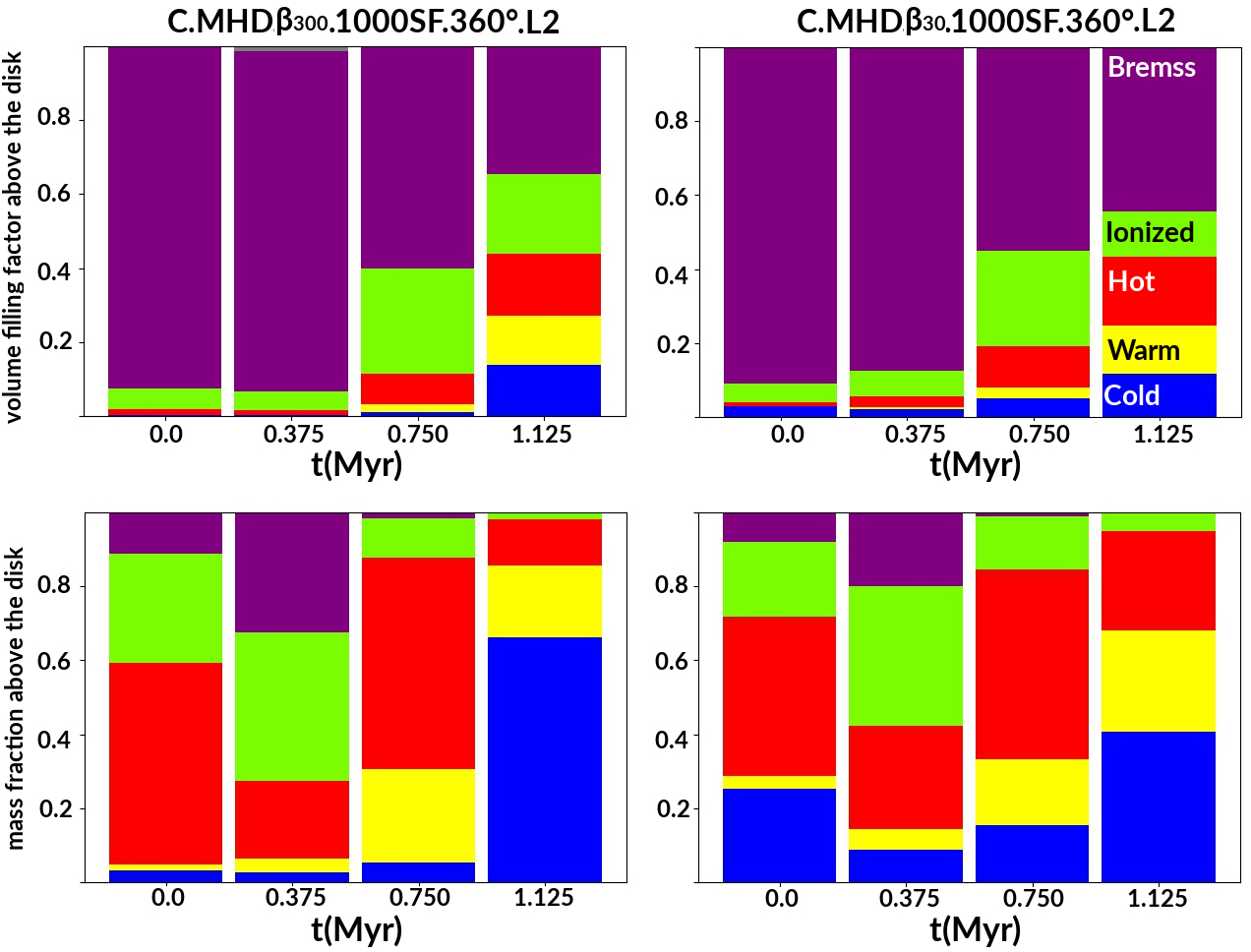}
\caption{\textbf{Volume filling factor} (top panels) and the associated mass fraction (bottom panels) in each temperature range  above the disk on both sides ($|z|> 200$pc) for  models differing only on the initial magnetic field strength: C.MHD$\beta_{300}$.1000SF.360°.L2 (left) and C.MHD$\beta_{30}$.1000SF.360°.L2 (right) shown in four different times:  t=0.0 Myr, t=0.375 Myr, t=0.750 Myr and t=1.125 Myr.
} \label{fT.beta300.vs.beta30}
\end{figure}

In Figure \ref{fT.beta300.vs.beta30}, we show the evolution of molecular gas above the disk in models with different magnetic field strength: $\beta$=300 and $\beta$=30, depicted in the left and right panels, respectively. It is worth noting that a stronger magnetic field presence enhances the formation of molecular gas. At $t=$0.75 Myr, the mass filling factor is approximately 15\% in contrast to the quasi-hydrodynamic model, which exhibits a value of around 5\%. However, by the end of the simulation, the magnetic field's increased intensity does not yield a significantly larger quantity of molecular gas. Consequently, we can infer that within the range of magnetic field values explored in this study, where thermal pressure surpasses magnetic pressure throughout the entire domain, the distribution of molecular gas remains unaffected.

\begin{figure}
\includegraphics[width=\columnwidth]{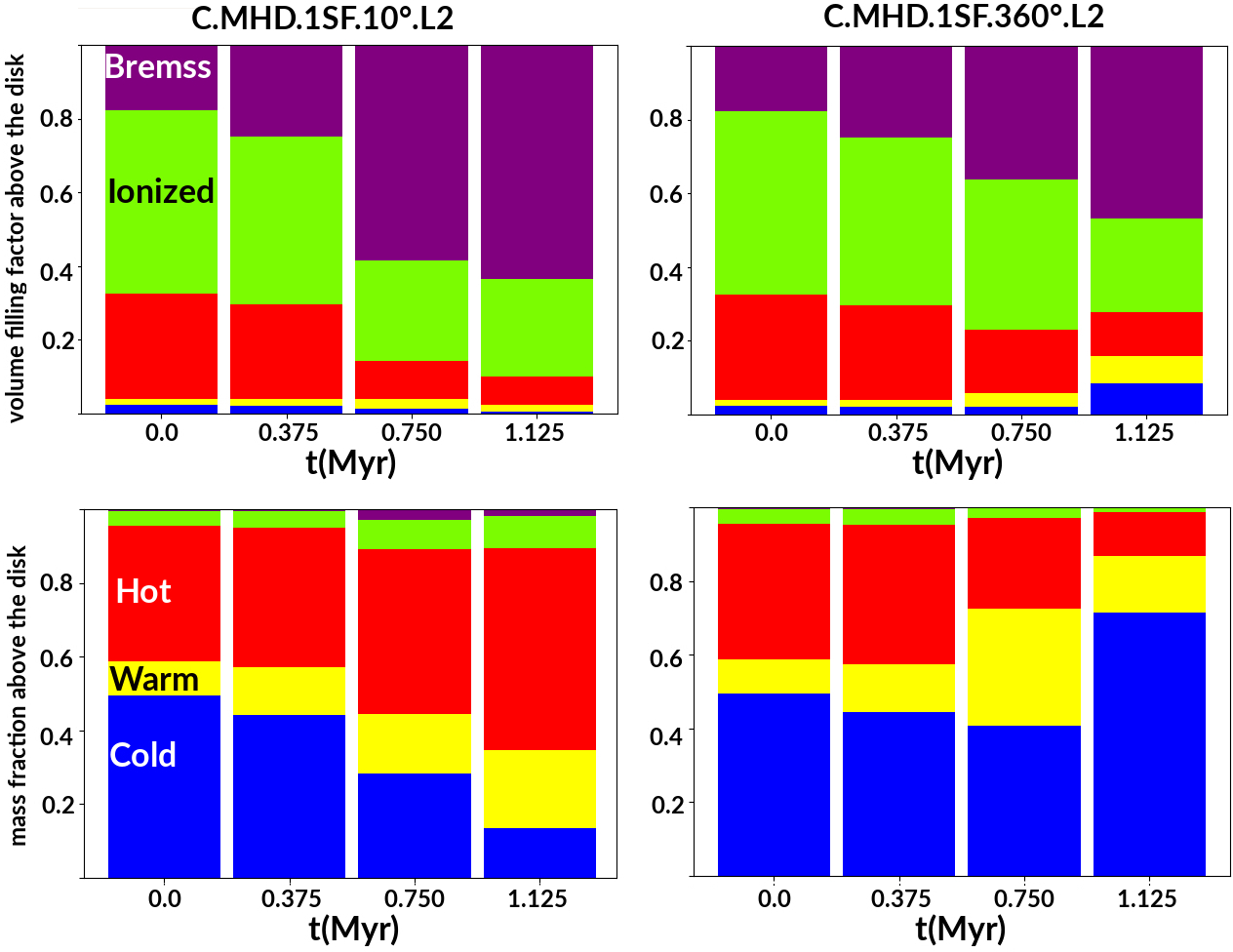}
\caption{\textbf{Volume filling factor} (top panels) and the associated mass fraction (bottom panels) in each temperature range above the disk on both sides ($|z|> 200$pc) for models with different AGN outflow opening angle, namely,   C.MHD.1SF.10°.L2 (left) and C.MHD.1SF.360°.L2 (right), shown for times  t=0.0 Myr, t=0.225 Myr, t=0.750 Myr and t=1.575 Myr. 
} \label{fT.Jet.vs.SPHER}
\end{figure}

\begin{figure}
\centering
\includegraphics[width=.8\columnwidth]{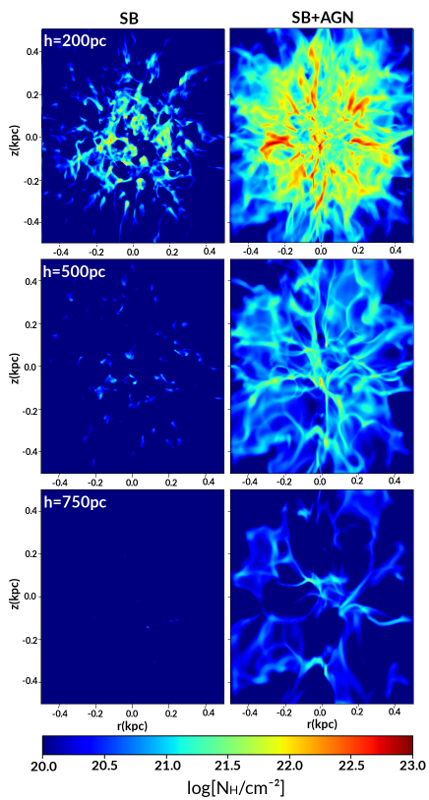}
\caption{\textbf{Face-on column density maps} outside the disk region at different heights:  $\pm$200 pc, $\pm$500 pc and $\pm$750 pc (top, middle and bottom panels, respectively) for a system with  SB-wind only (model C.MHD.100SF, left ) and a system with  SB+AGN winds (C.MHD.100SF.360°.L2, right), both at the snapshot t= 2.1 Myr. The SB+AGN system  drives  denser gas to outside the disk. At this time, the AGN wind has already terminated its activity, but its  passage, though not visible anymore, is the responsible for  transporting the dense gas above and below the disk.
}
\label{coldens.100SF.SB.vs.AGN}
\end{figure}

\begin{figure}
\centering
\includegraphics[width=.8\columnwidth]{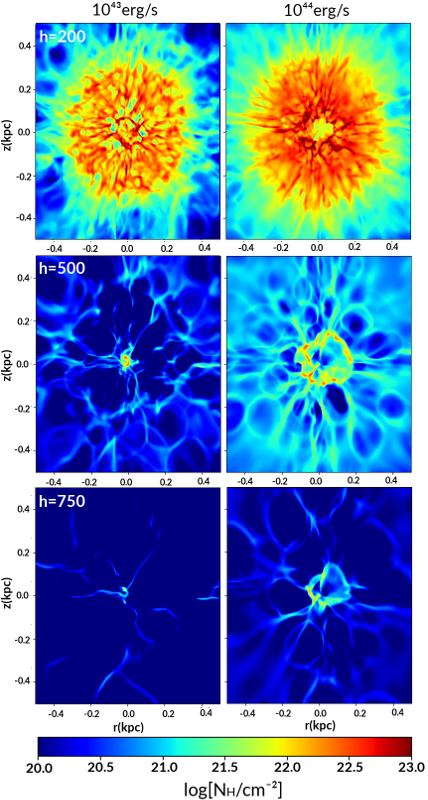}
\caption{\textbf{Face-on column density maps} outside the disk  at different heights:  $\pm$200 pc, $\pm$500 pc and $\pm$750 pc (top, middle and bottom panels, respectively) comparing models with  two different AGN wind luminosities: C.MHD.1000SF.360°.L2 and C.MHD.1000SF.360°.L3 both at the snapshot t= 2.1 Myr. The more luminous AGN wind can account for the transport of  more massive structures to outside the disk.
}
\label{coldens.43vs44}
\end{figure}

Figure \ref{fT.Jet.vs.SPHER} shows the filling factors  for two models with AGN wind opening angle 10° (left) and 360° (right), in different times. As time goes by, the fraction of cold gas  at high altitudes is clearly larger in the  360° model, as expected. The net amount of mass transported into the disk for the cold gas component in the 360° case is 
$\Delta \text{M}^{360^{\circ}}_{cold} \simeq 1.55\times10^7\text{M}_{\odot}$, while it decreases with time for the 10° case (by $\Delta \text{M}^{10^{\circ}}_{cold} \simeq -2.04\times10^6\text{M}_{\odot}$).
Figures \ref{coldens.100SF.SB.vs.AGN} and \ref{coldens.43vs44}  summarise well the results presented above, for the final stages of the systems.
Figure \ref{coldens.100SF.SB.vs.AGN} exhibits face-on column density maps obtained by excluding the disk within different ranges: $\pm$200 pc, $\pm$500 pc, and $\pm$750 pc, from the top to the bottom panels, respectively. The maps provide a comparative analysis of the impact of two systems: one featuring only the SB wind (model C.MHD.100SF, depicted on the left), and the other incorporating both the SB and AGN winds (model C.MHD.100SF.360°.L2, shown on the right). Essentially, these maps are like  a "tomography" of the region above and below the disk, showing the quantity and morphological characteristics of the denser (molecular) structures transported to varying heights above the disk.
It is evident that both models, with either the SB wind alone or the combination of SB and AGN winds, are capable of transporting dense gas to heights of approximately 1 kpc. However, the presence of the AGN wind noticeably enhances both the quantity and the maximum height of these structures.

Figure \ref{coldens.43vs44} presents similar column density maps, but for two models that differ only on the AGN luminosity. Clearly, the more luminous AGN wind can account for the transport of  more massive structures to the highest altitudes, outside  the disk.

\section{Discussion}
\label{sec:discussion}

The primary aim of this study was to determine the conditions under which an AGN wind can produce gas outflows that agree with observed data. To achieve this, the study sought to identify the essential mechanisms responsible for feedback and for generating the observed  molecular gas above the disk, as well as the initial conditions that must be considered in theoretical models to provide a justification for the observations. Key observables, such as the mass of the gas outflow, the mass of the extraplanar molecular gas, the SFR of the galaxy, and the luminosity of the AGN wind, were measured in a large sample of recent observations, providing a macroscopic yardstick for selecting the best and most realistic models from those under consideration here.

\subsection{Comparison with observations}\label{sec:observations}

In the upper part of Table \ref{table.cold.phase}, we summarize the main properties of the gas outflow obtained from our models with a special focus on the cold (molecular) gas component (50K-200K) and its kinematics. In the lower part of the table, for comparison, we depict the same quantities  for some selected nearby sources which show active star forming regions and AGN winds and which have, in general, similar values to those computed from our models \citep{Pereira-Santaella2018a,Fluetsch2019}, see also \citep{Pereira-Santaella2020, Pereira-Santaella2021,  Fluetsch2021, Lutz2019, Veilleux2020}.

\begin{figure*}
\includegraphics[width=\textwidth]{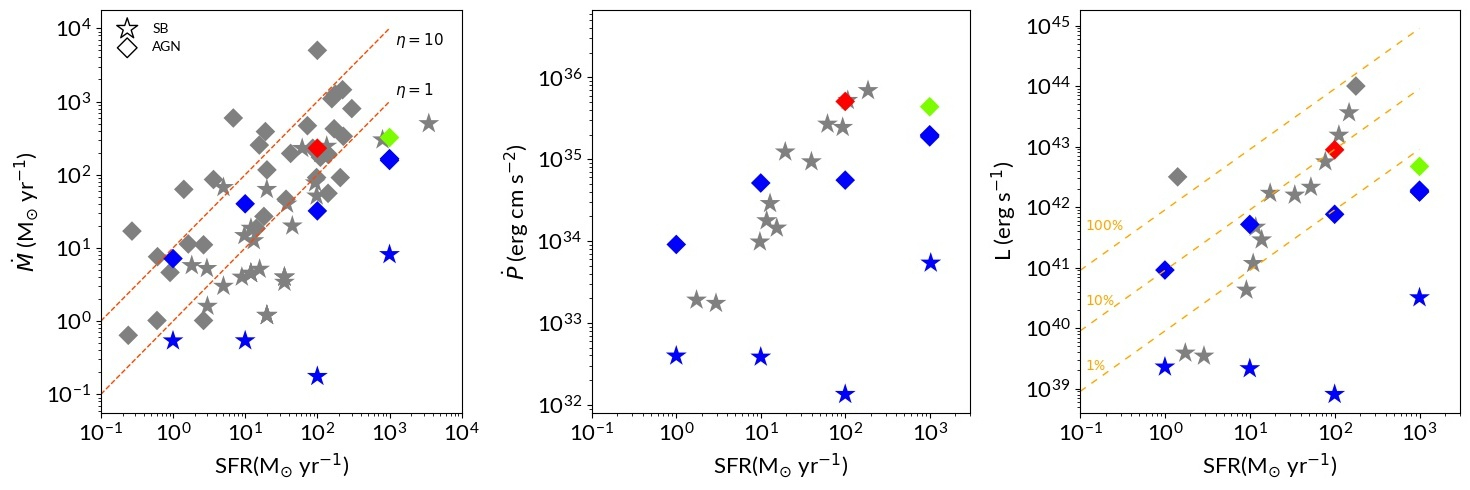}
\caption{Mass outflow rate vs.  SFR (left), outflow momentum rate vs.  SFR (middle), and outflow kinetic luminosity vs. SFR (right) of the cold (molecular) outflow component in our simulations compared to  observed sources. 
Diamonds indicate nuclei with outflows launched by an AGN, while stars represent nuclei with star-formation only. 
Gray symbols correspond to observed systems, namely, nearby ULIRGs, Sey galaxies, HII, and LINERS (\citealt{Garcia-Burillo2015}, \citealt{Jones2019}, \citealt{Pereira-Santaella2018a}).
The blue color stands for simulated models  with $L_{AGN}= L2= 2.35\cdot 10^{43}$ erg s$^{-1}$, while the red and green diamonds represent models with $L_{AGN}=L3=2.35\cdot10^{44}$erg s$^{-1}$.
The  dashed lines in the left panel give the mass loading factors, $\eta = \dot{M}/\text{SFR}$ of 1 and 10. 
The dashed  lines in the right panel give the percentage $a$  of the SFR  luminostiy ($L_{\text{SNe}}$) that is converted into outflow luminosity,   L$_{out}$ = a $\times$ L$_{\text{SNe}}$, with $a=100\%$, 10$\%$, and 1$\%$.}
\label{PS}
\end{figure*}

In Figure \ref{PS}, we plot these quantities, namely, the mass-loss rate $\dot{M}_{out}=M_{out}/t_{dyn}=M_{out}/(R_{out}/v_{out})$, the momentum rate 
$\dot{P}_{out}=\dot{M}_{out}v_{out}$, and the luminosity (mechanical power) $L_{out}=\dot{M}_{out}v_{out}/2$, where $R_{out}$ is defined as the outflow distance from the center, and we compare them with a larger sample of observed sources.
To calculate these quantities from our simulations, we compute  the outflow gas mass, $M_{out}$, with temperatures between $T=50K$ and $T=200K$, and vertical velocities greater than 100 km s$^{-1}$. Then, the outflow velocity  $v_{out}$ is calculated from the average of the vertical velocity of the cold structures. For most of the sources used for comparison here, the outflow distance $R_{out}$ ranges from $\sim$ 100 to  $\sim$ 400 pc, the velocity $v_{out}$ from 100 to 1000 km s$^{-1}$ and the corresponding outflow dynamical timescale, $t_{dyn}=R_{out}/v_{out}$ ranges from  10$^5$ to 10$^6$ yr.

The gray symbols in Figure \ref{PS}, represent observed sources, while the colored ones, the results from our models. Diamonds stand for systems with AGN winds, while stars, for systems with SF wind only.  The blue symbols stand for models   with $L_{AGN}= L2=2.35\cdot 10^{43}$ erg s$^{-1}$, while the red and green diamonds represent models with higher AGN wind luminosity, $L_{AGN}=L3=2.35\cdot10^{44}$erg s$^{-1}$ (and SFR = 100 and 1000 $M_{\odot}$ yr$^{-1}$, respectively).

The analysis of Figure \ref{PS} points to  four main results which are discussed below. 

The impact of the AGN wind helps to release the gas heated by star formation in the disk.
The  dashed lines in the left panel of Figure \ref{PS}, correspond to the mass loading factor, $\eta = \dot{M}/\text{SFR}$.
The leftmost blue diamonds in this  panel,  which represent  systems with AGN wind with luminosity $L_{AGN}= 2.35\cdot 10^{43}$ erg s$^{-1}$, and low SFR (SFR=1), have  a mass loading factor $\eta \sim 10$, while similar systems with SF wind only (blue stars), have $ \eta < 1$. This means that the  power generated by the SB wind alone is not enough to produce such a mass transport rate, as expected from our previous results.

Furthermore, we also see that maximum loading factors are obtained for the systems with AGN wind, and SFR=1 M$_{\odot}$ yr$^{-1}$.
For an  AGN wind  with same luminosity ($L2= 2.35\cdot 10^{43}$ erg s$^{-1}$), but with larger SFR, the mass loading factor decreases from $\eta \sim 10$ to  $\eta \sim 7$, for SFR=1 $M_{\odot}$ yr$^{-1}$ to 10 $M_{\odot}$ yr$^{-1}$, respectively, and   for systems with SFR=  100 and 1000 $M_{\odot}$ yr$^{-1}$,  $\eta <1$.
In other words, the AGN,  looses efficiency in carrying out mass to kpc distances in systems with higher SFR and column densities, suggesting that  for more massive systems, the mechanical power of the AGN should  be higher, as indicated by the observations. 
In fact,  the mass loading factor can be as high as $\eta \sim 4$ for a larger luminosity AGN, as indicated by the red  diamond in the left panel, which has a wind luminosity $L3=2.35\times10^{44}$erg s$^{-1}$.
The green diamond model, with same luminosity, but larger SFR = 1000 $M_{\odot}$ yr$^{-1}$ also has a higher mass loading than its lower luminosity counterpart (blue diamond), but still smaller than 1.  

The dashed  lines in the right panel of Figure \ref{PS} give the percentage $a$ of the SFR  luminositiy ($L_{\text{SNe}}$) that is converted into outflow luminosity,   $L_{out} = a \times L_{\text{SNe}}$. We see that  $L_{out}$ is approximately 10$\%$  $L_{\text{SNe}}$, when injected by  low SFR systems, and less than $1\%$ for  higher SFR systems. Similarly to the mass loading factor,  $L_{out}$ also increases with the AGN wind luminosity (as we see when comparing the lower luminosity blue diamonds with the higher luminosity red and green ones).

Finally, in  Figure \ref{PS} (left panel),  we can see that there are  several sources with the same SFR, but  different mass loss rates.
This could be due to different properties of their environments, characterizing 
 different stages specially  of the AGN evolution and activity.
 In fact, although we have plotted in the figure,  only  the maximum values of the mass loss rates obtained for the simulated models, we have seen in Figure \ref{disk.loss_sfr} that, as the AGN wind evolves with time in its  duty cycle of activity,  $\dot{M}$ also increases, suggesting a vertical motion upwards in the  $\dot{M}$ diagram of Figure \ref{PS}.
 
 {
 \setlength{\tabcolsep}{2.5pt}
 \begin{table*}
 \caption{Kinematic quantities of the cold outflow component ($T<$200K) from our simulations (upper part of the table): (1) name of the simulation; (2) total outflow mass, (3) total mass, (4) vertical outflow gas velocity, (5) dynamical time, $t_{\text{dyn}}=R_{\text{out}}/v_{\text{out}}$; (6) mass outflow rate, $\dot{M}_{\text{out}}=M_{\text{out}}v_{\text{out}}/R_{\text{out}}$; (7) depletion time, $t_{\text{dep}}=M_{\text{out}}/\dot{M}_{\text{out}}$; (8) momentum rate, $\dot{P}_{\text{out}}=\dot{M}_{\text{out}} v_{\text{out}}$; (9) kinetic luminosity, $L_{\text{out}}=\frac{1}{2}\dot{M}_{\text{out}}v_{\text{out}}^2$; (10) time of the maximum outflow rate;  (11) AGN luminosity; (12) AGN classification; and (13) star formation rate. The second and third parts of the table show the same quantities for observed sources with molecular outflows taken from \citet{Pereira-Santaella2018a} and \citet{Fluetsch2019}, respectively. Column (12) gives the observed source classification; the two values of SFR in column (13) for the molecular outflows from \citet{Pereira-Santaella2018a} were derived from the observed nuclear IR and non-thermal 1.4 GHz radio continuum luminosity, respectively.}
 \begin{tabular}{lcccccccccccc} \hline
 Models & log $M_{\text{out}}$ & log $M_{\text{tot}}$ & $v_{\text{out}}$ & log t$_{\text{dyn}}$ & $\dot{M}_{\text{out}}$ & log $t_{\text{dep}}$ & log $\dot{P}_{\text{out}}$ & log $L_{\text{out}}$ & $t_{\dot{M}}$ & log L$_{\text{AGN}}$ & Type & SFR \\ 
  & (M$_{\odot}$) & (M$_{\odot}$) & (km s$^{-1}$) & (yr) & (M$_{\odot}$ yr$^{-1}$) & (yr) & (gr cm s$^{-2}$) & (erg s$^{-1}$) & (Myr) & (erg s$^{-1}$) &  & (M$_{\odot}$ yr$^{-1}$) \\
  (1)&(2)&(3)&(4)&(5)&(6)&(7)&(8)&(9)&(10)&(11)&(12)&(13) \\ \hline
 C.MHD.1SF            &  6.03   &  8.50  &  117.4  &    6.22   &   0.7   &   8.69   &    32.68  &    39.45   &   1.06 &   -    & -  & 1      \\           
 C.MHD.1SF.360°       &  6.84   &  8.52  &  206.0  &    5.97   &   7.3   &   7.65   &    33.98  &    40.99   &   0.60 & 43,37  & -  & 1      \\           
 C.MHD.10SF           &  6.00   &  8.78  &  112.1  &    6.24   &   0.6   &   9.01   &    32.61  &    39.36   &   2.02 &   -    & -  & 10     \\           
 C.MHD.10SF.360°      &  7.60   &  8.85  &  198.5  &    6.00   &  40.0   &   7.24   &    34.70  &    41.70   &   1.42 & 43,37  & -  & 10     \\           
 C.MHD.100SF.         &  5.46   &  9.00  &  120.4  &    6.21   &   0.2   &   9.75   &    32.13  &    38.91   &   2.56 &   -    & -  & 100    \\           
 C.MHD.100SF.360°     &  7.36   &  9.06  &  273.6  &    5.85   &   31.6  &   7.55   &    34.74  &    41.88   &   1.80 & 43,37  & -  & 100    \\           
 C.MHD.1000SF         &  7.15   &  8.69  &  111.4  &    6.24   &   8.12  &   7.75   &    33.75  &    40.49   &   1.88 & 43,37  & -  & 1000   \\           
 C.MHD.1000SF.360°    &  8.23   &  8.86  &  187.0  &    6.01   & 162.2   &   6.64   &    35.29  &    42.26   &   2.47 & 43,37  & -  & 1000   \\           
 C.HD.1000SF.360°     &  8.21   &  8.86  &  184.6  &    6.01   & 155.0   &   6.64   &    35.26  &    42.22   &   2.32 & 43,37  & -  & 1000   \\ \hline    
\textbf{OBSERVATIONS}    &&&&&&&&&&&& \\     
 \text{Molecular Outflows} &&&&&&&&&&&& \\ 
 \citep{Pereira-Santaella2018a} \\ \hline 
 I12112 SW                  &  7.49   &  9.00  &  360.   &    6.3    &    17    &   7.9    &    34.39  &    41.74   & - & - &  ULIRG & 13/37  \\ 
 I12112 NE                  &  8.75   &  9.78  &  530.   &    6.1    &    398   &   7.2    &    36.13  &    43.55   & - & - &  ULIRG & 182/162  \\ 
 I14348 SW                  &  8.71   &  9.82  &  430.   &    6.4    &    251   &   7.5    &    35.79  &    43.13   & - & - &  ULIRG & 138/229  \\ 
 I14348 NE                  &  8.01   &  9.46  &  460.   &    6.1    &    80    &   7.5    &    35.38  &    42.75   & - & - &  ULIRG & 93/131  \\ 
 I22491 E                   &  8.07   &  9.54  &  400.   &    5.8    &    200   &   7.2    &    35.72  &    43.02   & - & - &  ULIRG & 110/58 \\ \hline
\text{Cold Molecular Outflows} &&&&&&&&&&&& \\ 
(\citealt{Fluetsch2019}) &&&&&&&&&&&& \\ \hline
NGC 1266               &   7,93   & -   &   177  &  -     &   11   &  8,19   & - & - & -  & 43,31  &  LINER  & 1,6  \\           
Circinus Galaxy       &   6,48   & -   &   150  &  -     &   1    &  9,31   & - & - & -  & 43,57  &  Sy2    & 0,6  \\           
M51                   &   6,61   & -   &   100  &  -     &   11   &  9,07   & - & - & -  & 43,79  &  Sy2    & 2,6  \\           
IRAS 15115 + 0208    &   8,82   & -   &   103  &  -     &   59   &  7.95   & - & - & -  & 43.49  &  H II   & 50,9 \\           
NGC 4418             &   7,90   & -   &   134  &  -     &   19   &  7.31   & - & - & -  & 43,81  &  Sy2    & 14,5 \\ \hline    
 \end{tabular}
 \end{table*}\label{table.cold.phase}
}


Table \ref{table.cold.phase} also gives values of the average velocity reached by clouds and filaments above the disk (cold structures with temperature between 50 K and 200 K with a vertical velocity of at least 100 km s$^{-1}$) obtained from our models. These velocities are between about 100 and 270 km s$^{-1}$, which are values similar to those observed in some sources such as NGC 1266, NGC 4418, IRAS 15115+0208, M51, and Circinus Galaxy,  which are highlighted in the bottom of the table and are classified as Seyferts or LINERS (\citealt{Fluetsch2019}). Although in our simulations we have identified few cold structures with maximum velocity of several 100 km/s (see e.g. Figure \ref{hst.nv.HD.vs.MHD}), the average  values obtained in the Table are lower than those observed for the sources classified as ULIRG, as reported in \cite{Pereira-Santaella2018a}. These average values also differ somewhat from those obtained by \cite{Mukherjee2018a, Mukherjee2018c}, where they detected dense clouds  dragged  along the jet with outflow velocities of $\sim$ 500 km s$^{-1}$.
Overall, our results suggest that only a very few dense structures can accelerate to  the observed velocities in ULIRGs, of about 500 km s$^{-1}$. However, we should remind that in our simulations the formation of cold structures in the disk is not imposed by the initial conditions, but instead,  is naturally driven by the combination of the star formation and radiative cooling in the ISM of the nuclear region of the galaxy, tested for different SFRs. So, what should be missing in our models?  First of all, we have not considered an important ingredient, namely, the presence of dust. The radiative cooling down to temperatures of dust formation may help to improve and prevent the evaporation of several cold structures. Moreover,  clouds may be further accelerated by the radiation pressure on dust relative to a hot, diffuse background, as shown in a series of numerical simulations (\citealt{Sharma_2013}, \citealt{Thompson2015}, \citealt{Yu2020}). Also, larger magnetic fields than those investigated in this work can  help to preserve a larger amount of cold structures from evaporation  (see e.g. \citealt{Farber2022MNRAS.510..551F}).
Finally,  higher resolution simulations could also allow for the formation  of thinner shock front shells that could, therefore,  accelerate cold filaments (that survive to evaporation) to high velocities, $in$ $situ$  at the high altitudes (\citealt{Viegas1992}). 
All these effects will be investigated in forthcoming work. 

Our simulations also indicate that the mechanism responsible for the  acceleration of cold clouds is not only due to the mechanical interaction between the jet and the interstellar medium. Of course, high SFR is able to produce a large fraction of molecular structures, and the action of the AGN wind is essential to remove these clumps from  the disk. Therefore, SF and AGN wind have to coexist to generate the observed molecular outflows. However, the velocities reached by the clouds are  in general lower than the escape velocity (of about 700 km s$^{-1}$), hence, eventually most of these structures will fall back onto the disk, as we see in the simulations. 
\subsection{Comparison with other simulated models}

The intense stellar feedback considered in our simulations is  the main source of turbulence in the ISM of the disk, but  further turbulence may also be driven transversely by the passage of the AGN wind. Both may contribute either to enhance SF or to quench it (e.g. \citealt{Schawinski2006}, \citealt{Mukherjee2018a}, \citealt{Barreto-Mota2021.503.5425B}), depending on features like the initial SFR (and thus the column density) and the AGN power, implying positive or negative feedback, respectively.  

\begin{figure*}
\includegraphics[width=\textwidth]{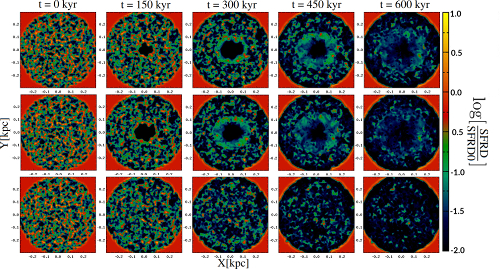}
\caption{Star formation rate surface density ($\text{SFRD}$) in the disk (defined as SFR per unit area) 
for the models C.MHD.100SF.360º with:  AGN wind luminosity L$_{\text{AGN}}=L2=2.35\cdot 10^{43}$erg s$^{-1}$ (top), L$_{\text{AGN}}=L3=2.35\cdot 10^{44}$erg s$^{-1}$ (middle), and with no AGN wind L$_{\text{AGN}}=0$ (bottom), for five different times of evolution. The empty region in the middle of the panels shows the imprint of the passage of the AGN-wind. The red ring in the edge of the disk is due to the very initial value of the SFR which is the same for the entire disk before the start of the SN explosions that lead to the clumpy environment in these models. }
\label{SFR100.NH23.C}
\end{figure*}

In order to compare our results with previous simulations (e.g., \citealt{Mukherjee2018a}), in Figure \ref{SFR100.NH23.C} we show the star formation rate surface density ($\text{SFRD}$) distribution in the disk relative to the initial value for the models C.MHD.100SF.360º for two AGN wind luminosities:  L$_2=2.35\cdot 10^{43}$erg s$^{-1}$ (top) and L$_3=2.35\cdot 10^{44}$erg s$^{-1}$ (middle), for different times of evolution from the active to inactive phases of the AGN duty cycle. 
In these diagrams, $\text{SFRD}$ is normalized to  $\text{SFRD0}$,  the star formation rate density at $t=0$.
SFR(t) is calculated from the Kenniccut-Schmidt law using eq. (3) of \cite{MelioliDalPino2015ApJ...812...90M} which relates this law to the column density of the gas, giving   $\text{SFRD}\sim 10^{-1}(n_{10}h_{\text{SB},200})^{1.4}$ M$_{\odot}$yr$^{-1}$kpc$^{-2}$, where $n_{10}$ is the density of the disk at $z=0$ in units of 10cm$^{-3}$ and $h_{\text{SB},200}$ is the height of the star burst region in units of 200pc. 
The cavity in the middle of the panels is due to the passage of the AGN-wind and it is larger for the AGN wind with larger luminosity (middle panel). 
The bottom diagram  shows how the same system behaves when there is no AGN outflow. We see that in all diagrams, there is a decrease in the SFRD with time caused by the passage of  both outflows. While  the SF wind acts more smoothly over all radii, the AGN wind causes the formation of a ring of enhanced star formation around the cavity during the active period ( $t \sim$ 150 kyr), which  disappears later on when the AGN wind  disappears  completely in the inactive phase ( $t \sim$ 600 kyr). This result can be  compared with that found by  \citealt{Mukherjee2018a} for a similar SFR $\sim$ 80 M$_{\odot}$yr$^{-1}$, but with collimated jet (see their model H). In their case, they  detect the formation of a stronger ring of enhanced SF around the AGN cavity, and decreasing SF in the outer regions.

\begin{figure*}
\includegraphics[width=\textwidth]{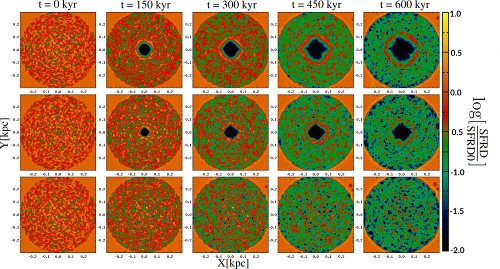}
\caption{The same as in Figure \ref{SFR100.NH23.C}, but for the  model with SFR 10 times larger,  C.MHD.1000SF.360°. In the top, the AGN wind luminosity is L$_{\text{AGN}}=L3=2.35\cdot 10^{44}$erg s$^{-1}$ (top),  in the middle L$_{\text{AGN}}=L2=2.35\cdot 10^{43}$erg s$^{-1}$, and in the bottom L$_{\text{AGN}}=0$.} \label{sigma.SFR1000.NH23}
\end{figure*}

In Figure \ref{sigma.SFR1000.NH23}, we have plotted the  same as in Figure \ref{SFR100.NH23.C},  but for the set of models  with SFR 10 times larger. In this case, the SFR density surface is clearly larger over the entire system, as expected. The ring of enhanced SF is also much stronger and  persists  even when the AGN  has completely disappeared in the  inactive phase, though less intense. 

\begin{figure*}
\includegraphics[width=\textwidth]{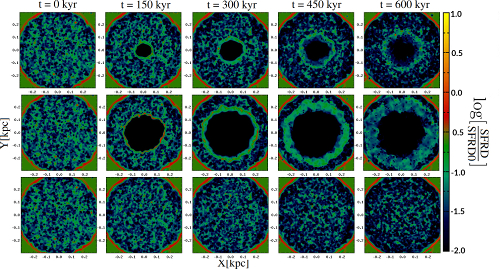}
\caption{The same as in Figure \ref{SFR100.NH23.C}, but for the models with SFR= 1  M$_{\odot}$ yr$^{-1}$ and distinct AGN wind opening angles (or none):  C.MHD.1SF.10º (top), C.MHD.1SF.360º (middle), and C.MHD.1SF (SB-only, bottom), in five different times of evolution.
Note that the thin red ring we see at 300 pc from the origin characterizes the initial SFRD in the system.
}
\label{sigma.SFR1.NH23}
\end{figure*}

In Figure \ref{sigma.SFR1.NH23}, we have plotted the evolution of the SFR surface density for models with a much smaller initial SFR (1 M$_{\odot}$yr$^{-1}$) and different AGN wind opening angle, namely, the conical  C.MHD.1SF.10º and  the spherical  C.MHD.1SF.360º models, both with luminosity L2, which are compared with the counterpart model with no AGN wind. 
In this case, we also clearly notice the formation of the strong ring around the cavity with enhanced SF as time evolves, reflecting a positive feedback of the AGN on SF. Its  counterpart model with opening angle 360º (bottom panel), shows  a much stronger effect with the AGN wind passage. The cavity is comparatively much bigger and the ring of enhanced SF around the wind is stronger.

Therefore, we conclude that the positive or negative feedback are highly sensitive to a non linear  combination of all the effects, that include the SF wind, the AGN  outflow opening angle and power,  the initial SFR (and column density), and the AGN wind phase of activity. The passage of the SF wind causes a decrease in the SFRD with time over the entire disk, and this effect is more pronounced for  larger SFR. The increase of the AGN power tends to enhance SF in a ring around the AGN wind that tends to disappear once the AGN wind faints, and this effect is also stronger for larger SFR. The wind opening angle also affects the SF feedback. A spherical injection creates a much larger cavity, but also a more enhanced ring of SF around the cavity.

\section{Conclusions}
\label {sec:Conclusions}

In this study we have presented three-dimensional MHD simulations of the core region of active galaxies at sub-kpc scales. Fiducial values of SF and SN rates were considered to allow for the natural formation of a turbulent and structured ISM where the AGN jet was injected. In this environment, a galactic wind driven by SF is also naturally produced, and the interactions and relative roles of both the AGN and the SF-driven outflows on the feedback processes were explored. The study also investigated the effects of initial conditions, including the intensity of mild magnetic fields, the intermittency of the AGN activity, the AGN outflow opening angle, the SFR in the disk, and the porosity generated in the ISM by stellar feedback. The underlying idea of this study was that the AGN wind alone would be unable to drive a massive gas outflow in the class of systems investigated, but it may have enough power to drag and accelerate clumps to very high velocities. The results of this study are summarized below.

\begin{enumerate}
\item The initial gas distribution determines the evolution of the system. In a smooth environment the AGN outflow expands faster, easily disrupting the disk, reaching the halo, and forming a low density, hot regular bipolar-shaped wind. On the other hand, in the case of the clumpy environment, an asymmetric, wiggling outflow develops on both hemispheres, propagating at slower velocity through the irregular structures that form in the thicker disk due to SF activity; 
\item After being continuously injected, the AGN wind completely removes the gas from the central region in approximately a few  $10^5$ yr, in all models investigated  (e.g. Figure \ref{mass.core.ULIRG-like}). Moreover, the high pressure of the AGN wind against the IS gas interrupts the gas fueling into the core which in turn, is responsible for the AGN outflow. This implies that the continuous injection of the AGN wind beyond about a few  $10^5$yr is no longer realistic. Without the pressure and energy injected by the AGN wind,  gas stellar feedback is able to resume   gas inflow that again refills the central region of the galaxy. Over time, the replenished gas reaches a maximum value that could eventually trigger a new phase of AGN wind activity. This evolution characterizes the duty cycle of the AGN. The entire cycle that includes the active, the remnant and the inactive phases lasts at most $\sim1.5$ Myr, being longer for larger SFR  and decreasing AGN wind power (Figure \ref{mass.core.ULIRG-like}); 
\item The impact of a collimated AGN wind (with an opening angle of approximately 0º - 10º) on the evolution of the galaxy disk in the nuclear region is significantly less pronounced compared to a spherical injection. In the case of a spherical injection, a larger working surface results in a greater amount of material being dragged to the halo, albeit at a slower expansion rate (see e.g. Figure \ref{TOTAL.JET.vs.SPHER});
\item  In the magnetized models, there is some deceleration of the AGN wind due to the impact into the magnetized gas which absorbs part of the kinetic energy at the terminal shock. However, for the magnetic fields we considered, with average values up to a few tens $\mu$G, as observed in most of the sources investigated here, we find no substantial effects in the macroscopic properties of the flow, or on  the evolution and formation of high-velocity  dense, cold structures;
\item The AGN wind looses efficiency in carrying out mass to kpc distances in systems with increasing  SFR
(and thus column densities). For instance, for an AGN wind luminostiy $\sim 10^{43}$erg s$^{-1}$, the total gas mass lost by the disk ($|z|<200pc$) varies from 10\% for  SFR=100-1000 M$_{\odot}$ yr$^{-1}$, to over 40$\%$ of the initial total mass for SFR$\leq$  10 M$_{\odot}$ yr$^{-1}$, but the increase in the AGN wind power by a factor 10 improves the mass loss rate by about a factor two.
On the other hand, models with  SB wind only, generally have smaller mass loss rates than their counterparts with AGN wind present (e.g. Figure \ref{disk.loss_sfr}). We find mass loss rates of the order of 50-250 M$_{\odot}$ yr$^{-1}$ in the first 2 Myr of evolution, for SFR in the range 1-1000 M$_{\odot}$ yr$^{-1}$  and luminosities in the range $10^{42-44}$erg s$^{-1}$, which are comparable to observed values in nearby Seyferts and ULIRGs;
\item Overall, we have found that the positive or negative feedback on star formation is highly sensitive to a non-linear combination of various effects, including the AGN outflow opening angle, power, and phase of activity, as well as the initial SFR (and column density). The passage of the SF wind results in a decrease in the SFR surface density over the entire disk with time. Increasing the AGN power enhances SF in a ring around the AGN wind, which disappears once the AGN wind weakens, and this effect is more pronounced for larger SFR. The wind opening angle also affects the SF feedback, with a spherical injection creating a larger cavity and an enhanced ring of SF around it than a collimated wind (see Figures \ref{SFR100.NH23.C}, \ref{sigma.SFR1000.NH23}, \ref{sigma.SFR1.NH23});
\item Finally, our models indicate that a higher SFR in the disk favors a better mixing of the IS matter with the AGN wind increasing the porosity of the disk and providing larger number of colder, denser structures (Figures \ref{DN.SFR.ON.AGN} to \ref{colden.SFR1-100}, and Figures \ref{append.ff.mf.NH24}-\ref{fT.bars.1000SB.vs.AGN}). The cold clouds and filaments, with temperatures between 50 K and 200 K, transported to distances above the disk between 200 pc and 1 kpc, reach average velocities of approximately 100 to 270 km s$^{-1}$. These velocities are comparable to those observed in several Seyferts and LINERS (e.g. reported in \cite{Fluetsch2019}). However, we have identified only very few cold structures with maximum velocities high enough to be compared to those observed in ULIRG systems, which are several hundred km s$^{-1}$, (e.g. \cite{Pereira-Santaella2018a}). Our average values differ from those obtained in other studies such as \cite{Mukherjee2018a, Mukherjee2018c}. It is important to note that in our simulations, the formation of cold structures in the disk is naturally driven by the combination of star formation and radiative cooling in the ISM of the nuclear region of the galaxy for different SFRs, instead of being imposed by initial conditions. In future work, we will include important missing ingredients such as the presence of dust and higher magnetic fields that may help to improve the formation and survival of high speed cold structures at high altitudes.
\end{enumerate}

\section*{Data availability}
The simulated data generated during this study  are available upon request to the authors.

\section*{Acknowledgements} The authors  acknowledge support   from the Brazilian Funding Agency FAPESP (grant 13/10559-5). WECB also acknowledges support  from the Brazilian Agency CNPq, and  EMdGDP from CNPq grant (308643/2017-8).  The  simulations presented in this work were performed in the cluster of the Group of Plasmas and High-Energy Astrophysics (GAPAE), acquired with support from  FAPESP (grant 2013/10559-5), and  the computing facilities of the Laboratory of Astroinformatics (IAG/USP, NAT/Unicsul), whose purchase was also made possible by FAPESP (grant 2009/54006-4) and the INCT-A. 

%
%
%
%
%
%
%


\bibliographystyle{mnras}
\bibliography{references} 








\bsp	
\label{lastpage}
\end{document}